\documentclass[tightenlines]{article}
\usepackage{jheppub}
\usepackage{amssymb}
\usepackage{amsmath}
\usepackage{epsfig}
\usepackage{tabularx}
\usepackage{inputenc}
\usepackage{url}
\usepackage{breqn}
\usepackage{pdflscape}
\usepackage{hyperref}
\usepackage{mathrsfs}
\usepackage{footmisc}
\usepackage{titlesec}
\usepackage{fancyhdr}
\usepackage{graphicx}
\usepackage{caption}
\usepackage{slashed}
\usepackage{wrapfig}
\usepackage{color}
\usepackage{tikz}
\usetikzlibrary{calc,trees,positioning,arrows,fit,shapes,calc}
\usepackage{appendix}

\tikzstyle{decision} = [diamond, draw, fill=blue!20, 
    text width=4.5em, text badly centered, node distance=3cm, inner sep=0pt]
\tikzstyle{block} = [rectangle, draw, fill=blue!20, 
    text width=7em, text centered, rounded corners, minimum height=4em]
\tikzstyle{line} = [draw, -latex']
\tikzstyle{cloud} = [draw, ellipse,fill=red!20, node distance=3cm,
    minimum height=2em]

\title{A global analysis strategy to resolve neutrino NSI degeneracies with scattering and oscillation data}

\author[a]{Bhaskar Dutta,}
\author[b]{Rafael F. Lang,}
\author[a]{Shu Liao,}
\author[c]{Samiran Sinha,}
\author[a]{Louis Strigari,}
\author[a]{and Adrian Thompson}
\emailAdd{thompson@physics.tamu.edu}

\affiliation[a]{Mitchell Institute for Fundamental Physics and Astronomy, Department of Physics  and Astronomy, Texas A$\&$M University,
College Station, TX 77843}
\affiliation[b]{Department of Physics and Astronomy, Purdue University, West Lafayette, IN 47907}
\affiliation[c]{Department of Statistics, Texas A$\&$M University, College Station, TX 77843}

\abstract{
Neutrino non-standard interactions (NSI) with the first generation of standard model fermions can span a parameter space of large dimension and exhibit degeneracies that cannot be broken by a single class of experiment. Oscillation experiments, together with neutrino scattering experiments, can merge their observations into a highly informational dataset to combat this problem. We consider combining neutrino-electron and neutrino-nucleus scattering data from the Borexino and COHERENT experiments, including a projection for the upcoming coherent neutrino scattering measurement at the CENNS-10 liquid argon detector. We extend the reach of these data sets over the NSI parameter space with projections for neutrino scattering at a future multi-ton scale dark matter detector and future oscillation measurements from atmospheric neutrinos at the Deep Underground Neutrino Experiment (DUNE). In order to perform this global analysis, we adopt a novel approach using the copula method, utilized to combine posterior information from different experiments with a large, generalized set of NSI parameters. We find that the contributions from DUNE and a dark matter detector to the Borexino and COHERENT fits can improve constraints on the electron and quark NSI parameters by up to a factor of 2 to 3, even when relatively many NSI parameters are left free to vary in the analysis.}


\begin{document}
\begin{flushright}
MI-TH-204
\end{flushright}

\maketitle

\section{Introduction}\label{sec:intro}
Non-standard neutrino interactions (NSI) are a popular effective field theory framework for exploring new physics beyond the standard model (BSM) in the neutrino sector~\cite{Wolfenstein:1977ue,Gavela:2008ra,Antusch:2008tz}. In the context of neutrino scattering experiments and neutrino oscillations, in the limit where any new gauge fields that mediate NSI are much heavier than the characteristic momentum transfer $q^2$, they are a convenient expression of the effective operators that arise in BSM extensions. NSI have been considered in many contexts and for a variety of neutral current and charged current operators. We limit the scope of this study to dimension-6 neutral-current (NC) vector NSI among the first-generation of SM fermions with real-valued couplings. They are described by the effective Lagrangian
\begin{align}
\label{eq:lagrangian}
    \mathscr{L}_{NSI} = -2\sqrt{2}G_F \sum_{f,\alpha,\beta}\bigg[ \epsilon^{f,L}_{\alpha\beta} (\bar{\nu}_\alpha \gamma^\mu P_L \nu_\beta)(\bar{f}\gamma_\mu P_L f)
    +\, \epsilon^{f,R}_{\alpha\beta} (\bar{\nu}_\alpha \gamma^\mu P_L \nu_\beta)(\bar{f}\gamma_\mu P_R f)\bigg].
\end{align}
Here the fermion indices are $f=e,u,d$ and flavor indices $\alpha,\beta=e,\mu,\tau$. $P_L$ and $P_R$ are the left and right projection operators, respectively. The effective dimension-6 operators in Eq.~\ref{eq:lagrangian} arise from some fundamental renormalizable theory~\cite{Dev:2019anc} where the NSI parameters $\epsilon^{f,L}_{\alpha\beta}$, $\epsilon^{f,R}_{\alpha\beta}$ are taken as proxies for the new propagators multiplied by couplings in the $M^2>>q^2$ limit. Experiments that are sensitive to different interaction channels and to different neutrino energies constrain fermion, flavor indicies, and projection operators. For example, solar neutrino experiments sensitive to neutrino-electron scattering are primarily sensitive to electron-type NSI, and also place the most stringent bounds on right-handed NSI.

In previous works that performed statistical analyses of NSI, it has been common practice to either consider a large family of NSI but only vary one or two of them at a time in the likelihood fit (Ref.~\cite{Giunti:2019xpr}, for example), or reparameterize the NSI down to a more phenomenological and pragmatically manageable subset based on model assumptions (for example, in Refs.~\cite{Gonzalez-Garcia:2013usa, Esteban:2018ppq, Esteban:2019lfo}). This is usually done for (i) the sake of model simplicity and (ii) computational limitations with regard to the dimensionality of the fit. However, in this work we are motivated to instead take an approach which is substantially more model-independent and generalized to more degrees of freedom.

Regarding (i) we note that a scenario in which more than two NSI are nonzero at once, albeit complex, have no good reason to be prohibited by nature. For those readers that may be interested in how NSI studies can guide model-building in the neutrino sector, a larger NSI parameter space is warranted to provide generalized constraints. Additionally, degeneracies among the NSI parameters arise due to transformations that leave the oscillation Hamiltonian and scattering cross sections invariant. The full space of these degeneracies as they show up in a likelihood analysis are not fully explored if only a small subset of NSI parameters are activated. Therefore, to explore this large-dimension scenario, we aim to perform a global analysis with all real-valued NSI in Eq.~\ref{eq:lagrangian} nonzero. A model-independent analysis of NSI of this breadth has not been performed to date. Although we will not consider complex-valued NSI, the phases $\phi^{f,P}_{\alpha\beta}$ in the decomposition $\epsilon^{f,P}_{\alpha\beta} = |\epsilon^{f,P}_{\alpha\beta}| e^{i\phi^{f,P}_{\alpha\beta}}$ exhibit more pronounced degeneracies with the PMNS parameters such as the CP-violating phase $\delta_{CP}$ and other vacuum parameters~\cite{Esteban:2019lfo,Masud:2015xva}, which we consider fixed in this work. We will restrict our focus to degeneracies between real NSI and reserve a maximally-general treatment of complex-valued parameters and their degeneracies for a future analysis.

To address issue (ii), we have developed a new statistical technique by virtue of divide-and-conquer which allows one to perform a large-dimensional analysis with a variety of experimental data that are sensitive to different linear combinations of the NSI parameters. The tool in question which allows us to pursue this study without technological barriers is the \textit{copula}, a statistical object popularized in other data-driven fields but which is quite novel to particle physics. We will discuss this technique in detail in Section~\ref{sec:priorflow}.

While working in the context of a many-parameter NSI study, the degeneracies that present themselves in physical observables motivate a specific combination of experimental data that can break such degeneracies. While looking forward to the plentiful source of neutrino oscillation data at DUNE, we also raise awareness that DUNE's excellent projected sensitivity will only be indirectly sensitive to the electron, $u$ and $d$ quark NSI via their linear combination that enters into the matter potential of the oscillation Hamiltonian. We therefore recognize the need to augment DUNE's future oscillation measurements with neutrino scattering data, namely those from the COHERENT and Borexino experiments, which have more direct access to these NSI. Additionally, as the next generation of multi-ton scale dark matter detectors will be sensitive to neutrino interactions, they also provide a means to study NSI. In particular, natural neutrino sources such as the solar and atmospheric neutrino fluxes contain $\tau$-flavor neutrinos which can complement the $\nu_\tau$-deficient neutrino source at COHERENT. In this work we envision a unified experimental dataset comprised of neutrino oscillation and scattering data at COHERENT and Borexino, joined with future projections for DUNE and a ton-scale dark matter detector, to carry out a generic, multi-dimensional NSI analysis.

The paper is organized as follows. In Section~\ref{sec:degen} we break down the varieties of degeneracies among the NSI parameters and discuss how these degeneracies may complement each other in a global analysis. In Section~\ref{sec:priorflow} we demonstrate our global analysis strategy and in Section~\ref{sec:methods} we briefly outline our analysis methods for each experiment under consideration. Finally in Section~\ref{sec:results} we present and discuss the posterior distributions of all the NSI parameters included in the analysis and in Section~\ref{sec:conclusion} we conclude.

\section{Degeneracies}\label{sec:degen}
The phenomenology of neutrinos scattering with SM fermions in the first generation exhibit several experimental degeneracies, or transformations in the NSI parameter space that leave a physical observable such as a cross-section or a Hamiltonian invariant. These degeneracies leave their footprint directly in the likelihood profiles derived from scattering and oscillation data due to the way the NSI parameters enter into the fundamental observables; therefore, it is important to understand the degeneracy structures in order to know how they can be broken. We outline two such classes of degeneracies between NSI parameters that present themselves in neutrino oscillation and scattering data - those that exhibit degeneracy between NSI parameters of different fermion index $f$, and those between different flavor indices $\alpha\beta$. Many of the degeneracies we will discuss have already been derived and discussed before in the literature, but we include them here to have a complete motivation of the subsequent analysis.

\subsection{``Fermion'' Degeneracies}\label{sec:fermion_degen}
The first class of degeneracies concerns the ability for an experiment to distinguish NSI between different SM fermions, $f$ and $f^\prime$, manifested between $\epsilon^{f,V}_{\alpha\beta}$ and $\epsilon^{f^\prime,V}_{\alpha\beta}$, for example. This type of degeneracy manifests itself differently within three important classes of interactions, namely neutrino oscillations, neutrino-nucleus scattering, and neutrino-electron scattering.

\subsubsection{Oscillation Experiments}
An experiment measuring neutrino oscillations through the Earth has direct sensitivity to the oscillation Hamiltonian and its NSI contribution to the matter potential;
\begin{equation}
    H_{\alpha\beta} = \dfrac{1}{2 E_\nu} (U M U^\dagger)_{\alpha\beta} + V^{Matter}_{\alpha\beta}
\label{eq:hamiltonian}
\end{equation}
taking 3 flavors $\alpha$,$\beta = e$, $\mu$, $\tau$. The first term, dependent on the neutrino energy $E_\nu$, controls flavor oscillations in vacuum. It contains the mixing angles $\theta_{12}$, $\theta_{13}$, and $\theta_{23}$ within the PMNS mixing matrix $U$ and the neutrino mass splittings in $M = \text{diag}[0,\Delta m_{21}^2, \Delta m_{31}^2]$, which we have taken to be in normal heirarchy ($\Delta m_{31}^2 > \Delta m_{21}^2$)\footnote{In this work we fix $\theta_{12} = 0.576$, $\theta_{13} = 0.148$, $\theta_{23} = 0.722$, $\Delta m_{21}^2 = 7.37\cdot 10^{-17}$ MeV$^2$, and $\Delta m_{31}^2 = 2.54\cdot 10^{-15}$~MeV$^2$.}. Neutrino NSI are contained in the matter potential $V$ which is a function of the coordinate $x$; 
\begin{equation}
V^{Matter}_{\alpha\beta}= \sqrt{2}G_F \Big[ n_e(x)(\delta_{\alpha e}\delta_{e\beta} + \epsilon^{e,V}_{\alpha \beta})+ n_u(x)\epsilon^{u,V}_{\alpha \beta}+ n_d(x)\epsilon^{d,V}_{\alpha \beta} \Big]
\label{eq:matter_pot}
\end{equation}
where $\epsilon^{f,V}_{\alpha\beta} \equiv \epsilon^{f,L}_{\alpha\beta} + \epsilon^{f,R}_{\alpha\beta}$. This potential term supports Wolfenstein oscillations in matter~\cite{Wolfenstein:1977ue}, as well as richer phenomena, such as the Mikheyev-Smirnov-Wolfenstein (MSW) effect~\cite{Smirnov:2004zv, Mikheev:1986gs} and parametric resonances of neutrino oscillations through discretely varying matter densities~\cite{Akhmedov:1998ui}, e.g., the Earth. It is dependent on the electron, up-quark and down-quark number densities as well as the NSI parameters. The NSI terms can be expressed as a matrix containing all the flavor $\alpha\beta$ vertices;
\begin{equation}
    \epsilon^{f,V} = \begin{pmatrix} 
\epsilon_{ee}^{f,V} & \epsilon_{e\mu}^{f,V}& \epsilon_{e\tau}^{f,V}\\
\epsilon_{e\mu}^{f,V} & \epsilon_{\mu\mu}^{f,V} & \epsilon_{\mu\tau}^{f,V} \\
\epsilon_{e\tau}^{f,V} & \epsilon_{\mu\tau}^{f,V}  & \epsilon_{\tau\tau}^{f,V}
\end{pmatrix}
\end{equation}

The electron, up, and down NSI enter into the oscillation Hamiltonian in linear combination (Eq.~\ref{eq:matter_pot}), and upon diagonalization it is this linear combination that enters into the survival and transition probabilities as physical observables. One may refer to~\cite{Parke:2019vbs} for the treatment of neutrino oscillation probabilities, and more recently~\cite{Denton:2019ovn} for the exact amplitudes. To simplify things, the electron number density can be factored out and since $n_{u} \approx n_{d} \approx  3 n_e$ in the Earth to good approximation, any experiment that measures the matter potential effects of neutrino oscillation is only sensitive to the sum. We therefore define a \textit{phenomenological} NSI parameter;
\begin{equation}
\label{eq:osc_fermion_degen}
    \epsilon_{\alpha\beta}^{O} \equiv \epsilon^{e,V}_{\alpha\beta} + 3 \epsilon^{u,V}_{\alpha\beta} + 3\epsilon^{d,V}_{\alpha\beta}
\end{equation}
This is the NSI observable for experiments measuring NSI in oscillations through the Earth. It defines a plane of solutions in $(\epsilon^{e,V}_{\alpha\beta}, \epsilon^{u,V}_{\alpha\beta}, \epsilon^{d,V}_{\alpha\beta})$ space; unless two out of the three terms are fixed, oscillation experiments have a three-fold degeneracy in their sensitivity to the $e$, $u$, and $d$ NSI.

\subsubsection{CE$\nu$NS Experiments}
Neutrinos interact coherently with nuclei if their transferred momentum $q$ satisfies $q r_n <<1$ for a nuclear radius $r_n$. This process is described via the Coherent Elastic Neutrino-Nucleus Scattering (CE$\nu$NS) mechanism \cite{PhysRevD.9.1389, Scholberg:2005qs}. In this energy range (typically for $E_\nu$ as high as 100 MeV for most nuclei), the first term in Eq.~\ref{eq:hamiltonian} becomes large leaving the matter potential as a subdominant effect and relegating new physics observables to the CE$\nu$NS cross-section. The cross-section is given by
\begin{equation}
    \dfrac{d\sigma}{d E_r} = \dfrac{G_F^2 Q_V^2 m_N}{2 \pi} \bigg(1 - \dfrac{m_N E_r}{E_\nu^2} + \bigg(1 - \dfrac{E_r}{E_\nu} \bigg)^2 \bigg) F(q^2)
\label{eq:cevns}
\end{equation}
where we traditionally take the Helm parameterization of the form factor $F(q^2)$ with a neutron skin radius $r_n = 5.5$ fm~\cite{Cadeddu:2017etk, PhysRev.104.1466}.
The $Q_V^2$ factor ordinarily contains the SM charges, but with the presence of NSI, potentially allowing flavor changing processes such as $\nu_\alpha + N \rightarrow \nu_\beta + N$, it is modified to
\begin{align}
Q_V^2 &= 4\bigg[Z \bigg(\frac{1}{2} - 2s_w^2 + 2\epsilon^u_{\alpha\alpha} + \epsilon^d_{\alpha\alpha}\bigg) + N \bigg(\epsilon^u_{\alpha\alpha} + 2\epsilon^d_{\alpha\alpha} -\frac{1}{2} \bigg)\bigg]^2 \nonumber \\
&+ 4\sum_{\beta \neq \alpha} \bigg| Z(2\epsilon^u_{\alpha\beta}+\epsilon^d_{\alpha\beta}) + N(\epsilon^u_{\alpha\beta}+2\epsilon^d_{\alpha\beta})  \bigg|^2
\end{align}
where $s_w = \sin\theta_w$ is the sine of the Weinberg angle and we have taken the initial state flavor $\alpha$ and summed over final state flavors. $Z$ and $N$ are the proton and neutron numbers of the target nucleus, respectively. A linear combination of the up and down vector NSI, which we denote $\epsilon_{\alpha\beta}^{N}$, can then be factored out such that $Q_V$ is a function of a single NSI parameter, which we denote by $\epsilon_{\alpha\beta}^{N}$;
\begin{equation}
\epsilon_{\alpha\beta}^{N} \equiv \epsilon^{u,V}_{\alpha\beta} + \frac{(2N+Z)}{(2Z+N)}\epsilon^{d,V}_{\alpha\beta}
\end{equation}
giving us our second \textit{phenomenological} NSI parameter. This time the linear combination of up and down pieces is responsible for degeneracy between $\epsilon^{u,V}_{\alpha\beta}$ and $\epsilon^{d,V}_{\alpha\beta}$ NSI in CE$\nu$NS experiments. Upon squaring $Q_V$ and setting $\epsilon_{\alpha\beta}^{N}$ equal to a constant reveals a set of solutions in the ($\epsilon^{u,V}_{\alpha\beta},\epsilon^{d,V}_{\alpha\beta}$) plane. For example, consider a single flavor-diagonal NSI $\epsilon^N_{\alpha\alpha}$. Setting $Q_V^2 (\epsilon^N_{\alpha\alpha}) = Q_V^2 (0)$ to find the solutions degenerate with the SM gives
\begin{equation}
    (\epsilon^N_{\alpha\alpha})^2 - \dfrac{Q_V}{2Z+N}\epsilon^N_{\alpha\alpha} = 0
\end{equation}
giving rise to two lines of solutions in the ($\epsilon^{u,V}_{\alpha\beta},\epsilon^{d,V}_{\alpha\beta}$) plane;
\begin{equation}
\label{eq:cevns_fermion_degen}
\left\{
    \begin{aligned}
        \epsilon^{u,V}_{\alpha\alpha} + \dfrac{2N+Z}{2Z+N}\epsilon^{d,V}_{\alpha\alpha} - \dfrac{Q_V}{2Z+N} &= 0 \quad\\
        \epsilon^{u,V}_{\alpha\alpha} + \dfrac{2N+Z}{2Z+N}\epsilon^{d,V}_{\alpha\alpha} &= 0
    \end{aligned}
\right\}
\end{equation}
The slope of these lines is given by the ratio $(2N+Z)/(2Z+N)$ which varies depending on the detector material; for instance, $^{126}_{54}$Xe has a ratio of 1.1 while $^{40}_{18}$Ar has 1.07. Therefore using CE$\nu$NS data from multiple detectors of different materials can have complementary likelihood profiles that can help to break this type of degeneracy~\cite{Barranco:2005yy,Dent:2017mpr, Scholberg:2005qs}.

\subsubsection{Elastic Neutrino-Electron Scattering (E$\nu$ES) Experiments}
Lastly, we consider the elastic neutrino-electron scattering (which we will refer to as E$\nu$ES) cross-section, measured from the Solar neutrino flux, for example, contains a similar charge structure when NSI are included. It may also permit flavor-changing scattering processes $\nu_\alpha + e^- \rightarrow \nu_\beta + e^-$;
\begin{align}
\dfrac{d \sigma_{\alpha\beta}}{dE_r} = 2\dfrac{G_F^2 m_e}{\pi} \bigg[& (\delta_{e\alpha}\delta_{e\beta} + \delta_{\alpha\beta}g_L + \epsilon^{e,L}_{\alpha\beta})^2 + (\delta_{\alpha\beta}g_R + \epsilon^{e,R}_{\alpha\beta})^2 \bigg(1 - \frac{E_r}{E_\nu}\bigg)^2 \nonumber \\
&- (\delta_{\alpha\beta}g_L + \epsilon^{e,L}_{\alpha\beta})(\delta_{\alpha\beta}g_R + \epsilon^{e,R}_{\alpha\beta}) \dfrac{m_e E_r}{E_\nu^2} \bigg]
\label{eq:eves}
\end{align}
where $g_L = \sin^2 \theta_w - \frac{1}{2}$ and $g_R = \sin^2 \theta_w$. The $\delta_{e\alpha}\delta_{e\beta}$ term encodes the charged-current enhancement for $\nu_\alpha + e^- \rightarrow \nu_\beta + e^-$ scattering, and the other Kronecker delta terms take care of removing the SM charges when the process is flavor-changing. Unlike the CE$\nu$NS differential cross-section, there is only one fermion index appearing in Eq.~\ref{eq:eves} and since $\epsilon^{e,R}_{\alpha\beta}$ and $\epsilon^{e,L}_{\alpha\beta}$ appear next to terms of different energy dependence, it is in principle possible to disentangle them in a multiparameter fit to data. For this reason, we simply identify the \textit{phenomenological} NSI with the physical NSI;
\begin{align}
\epsilon^{E,R}_{\alpha\beta} &\equiv \epsilon^{e,R}_{\alpha\beta} \nonumber \\
\epsilon^{E,L}_{\alpha\beta} &\equiv \epsilon^{e,L}_{\alpha\beta}
\end{align}
However, the way these NSI appear in the cross-section gives rise to a more complicated degeneracy structure than in CE$\nu$NS and oscillation experiments, since the left and right chiral NSI parameters do not simply factor out of the energy-dependent terms in Eq.~\ref{eq:eves}. We can visualize the degenerate features of this cross-section by equating the NSI cross-section with the SM one and match like-terms in the $E_r$ expansion;
\begin{equation}
\left\{
\begin{aligned}
(\delta_{e\alpha}\delta_{e\beta} + \delta_{\alpha\beta}g_L + \epsilon^{e,L}_{\alpha\beta})^2 + (\delta_{\alpha\beta}g_R + \epsilon^{e,R}_{\alpha\beta})^2 &= (\delta_{e\alpha}\delta_{e\beta} + \delta_{\alpha\beta}g_L)^2 + \delta_{\alpha\beta}g_R^2 \\
2(\delta_{\alpha\beta}g_R + \epsilon^{e,R}_{\alpha\beta})^2 + \dfrac{m_e}{E_\nu}(\delta_{\alpha\beta}g_R + \epsilon^{e,R}_{\alpha\beta})(\delta_{\alpha\beta}g_L + \epsilon^{e,L}_{\alpha\beta}) &= \delta_{\alpha\beta}(2g_R^2 + \dfrac{m_e}{E_\nu}g_R g_L) \\
(\delta_{\alpha\beta}g_R + \epsilon^{e,R}_{\alpha\beta})^2 &= \delta_{\alpha\beta}g_R^2
\end{aligned}
\right\}
\end{equation}

This system of equations has one or two solutions depending on the presence of the $\delta_{e\alpha}\delta_{e\beta}$ term (and in the two solution case, the larger of these solutions may be disallowed by existing constraints). Note that as $m_e / E_\nu \to 0$, the number of equations reduces to 2 and the number of solutions rises to 4, therefore encouraging the measurement of this cross-section at relatively lower neutrino energies, for example, the $^7$Be solar neutrino flux ($E_\nu \approx 0.86$ MeV).
\begin{equation}
\label{eq:eves_fermion_degen}
\left\{
    \begin{aligned}
    \epsilon^{e,L}_{\alpha\beta} = \epsilon^{e,R}_{\alpha\beta} &= 0 \\
    \epsilon^{e,L}_{\alpha\beta} = -2g_L, \epsilon^{e,R}_{\alpha\beta} &= -2g_R
    \end{aligned}
    \right\}
\end{equation}
For $\alpha = \beta = e$ only the top line of Eq.~\ref{eq:eves_fermion_degen} is a solution, but for all other $\alpha\beta$ pairs both solutions exist, implying at most two solutions for $\epsilon^{e,V}_{\alpha\beta}$. The Eqs.~\ref{eq:eves_fermion_degen} are shown in Appendix \ref{app:degen} for both of the aforementioned cases. We will also examine what happens in the case that more than one NSI flavor index is nonzero at once in the following section. These degeneracy structures have been studied in more detail in the context of the DUNE near detector in~\cite{Bischer:2018zcz}.\\

\begin{figure}[h]
 \centering
 \includegraphics[width=0.5\textwidth]{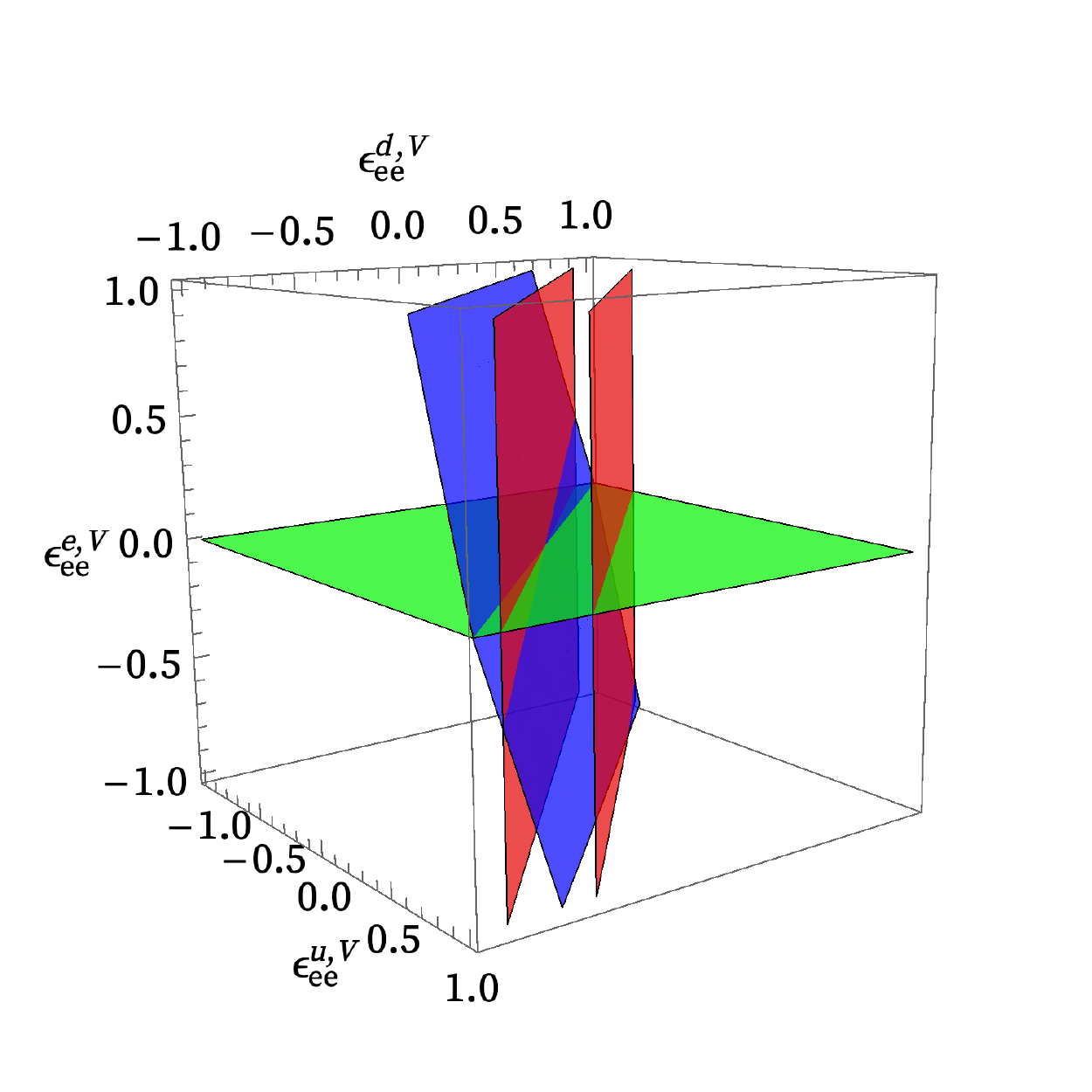}
\caption{\label{fig:fermion_planes} The overlapping plane solutions for CE$\nu$NS (red), E$\nu$ES (green), and oscillation (blue) experiments for a single flavor index $ee$. In this scenario, the intersection point where all three sets of solutions hold is at the origin (SM).}
\end{figure}

Collecting these three fermion degeneracies together in the simple case of a single flavor index activated, we show the overlapping solutions which are degenerate with the SM in $(\epsilon^{u,V}_{ee},\epsilon^{d,V}_{ee},\epsilon^{e,V}_{ee})$ space in Figure~\ref{fig:fermion_planes}. These plane solutions correspond to Eqs.~\ref{eq:osc_fermion_degen}, \ref{eq:cevns_fermion_degen}, and \ref{eq:eves_fermion_degen}. The point at which all solutions simultaneously intersect, i.e., the maximum likelihood point for a likelihood function defined over the combination of oscillation, CE$\nu$NS, and E$\nu$ES data, is at the origin. The important implication here is that the degeracies among triads of $(\epsilon^{u,V}_{\alpha\beta},\epsilon^{d,V}_{\alpha\beta},\epsilon^{e,V}_{\alpha\beta})$ NSI parameters can be broken by combining the three aforementioned experimental classes.

\subsection{``Flavor'' Degeneracies}
In addition to providing control over degeneracies between $e$, $u$, and $d$ NSI, there is another class of degeneracies present to which oscillation and scattering experiments can conspire to resolve, existing between NSI of different flavor vertices, for example, between $\epsilon^{f,V}_{\alpha\beta}$ and $\epsilon^{f,V}_{\gamma\delta}$.\\

\subsubsection{Oscillation Experiments}
Any Hamiltonian is invariant under a translation proportional to the identity matrix such as $H_{\alpha\beta} \to H_{\alpha\beta} - C\delta_{\alpha\beta}$. Supposing that we take all three flavor-diagonal NSI to be nonzero, we can conventionally take $C =  \epsilon^O_{\mu\mu}$ and see that the transformations 

\begin{align}
\epsilon^O_{ee} \to \epsilon^O_{ee} - \epsilon^O_{\mu\mu} \nonumber \\
\epsilon^O_{\tau\tau} \to \epsilon^O_{\tau\tau} - \epsilon^O_{\mu\mu}
\end{align}
leave $H$ invariant. If one were to measure the $ee$ and $\tau\tau$ components of the matter potential (Eq.~\ref{eq:matter_pot}), the solutions degenerate with the standard model values would yield the equations for two planes;

\begin{align}
\epsilon^O_{ee} - \epsilon^O_{\mu\mu} = 0 \nonumber \\
\epsilon^O_{\tau\tau} - \epsilon^O_{\mu\mu} = 0
\end{align}
Subtracting the two equations verifies the existence of the  $\epsilon^O_{ee} - \epsilon^O_{\tau\tau} = 0$ degenerate plane as well. Usually, analyses that are exclusively sensitive to NSI through oscillation data will take the reparameterization $\epsilon_{ee}^{O} \to \epsilon^O_{ee} - \epsilon^O_{\mu\mu}$ and $\epsilon_{\tau\tau}^{O} \to \epsilon^O_{\tau\tau} - \epsilon^O_{\mu\mu}$ to transform away the degeneracy.

Another source of degeneracy arises from the CPT symmetry $H \to -(H)^*$. This symmetry is manifested in the matter potential as
\begin{align}
    \epsilon^O_{ee} - \epsilon^O_{\mu\mu} &\rightarrow -(\epsilon^O_{ee} - \epsilon^O_{\mu\mu}) - 2  \nonumber \\
\epsilon^O_{\tau\tau} - \epsilon^O_{\mu\mu} &\rightarrow -(\epsilon^O_{\tau\tau} - \epsilon^O_{\mu\mu}) 
\end{align}
and an additional transformation in the off-diagonal NSI parameters, $\epsilon^O_{\alpha\beta} \rightarrow -(\epsilon^O_{\alpha\beta})^*$, which we do not consider in this work for limiting ourselves to real-valued NSI. These transformations become especially relevant when one also considers the mass-ordering parameters, mixing angles and phases in the vacuum part of the Hamiltonian to vary alongside NSI, giving rise to generalized mass-ordering and mixing angle degeneracies~\cite{Coloma:2016gei, Deepthi:2016erc, Coloma:2017egw, Gonzalez-Garcia:2013usa, GonzalezGarcia:2011my, Miranda:2004nb}.

\subsubsection{CE$\nu$NS Experiments}
Allowing more than one NSI of different flavor indices $\alpha\beta$ within the $Q_V^2$ factor of the CE$\nu$NS cross-section gives rise to generalizations of the plane solutions outlined in the previous section. To see this, once again we equate the NSI-modified $Q_V^2$ with the SM $Q_V^2$ and find all the solutions. For example, taking the incoming neutrino flavor as $e$, we find the equation
\begin{equation}
\label{eq:cevns_flav_degen}
    Q_V^2 = \bigg[-Q_V + 2(2Z+N) \epsilon^N_{ee} \bigg]^2
    + 4(2Z+N)^2 \bigg[(\epsilon^N_{e\mu})^2 + (\epsilon^N_{e\tau})^2 \bigg]
\end{equation}
Expanding Eq.~\ref{eq:cevns_flav_degen} into the $u$ and $d$ components yields the equation of a hyperellipse in 6 dimensions, but depending on the number of NSI included, this hyperellipse breaks down into simpler solution sets. For example, in the case that we have $\epsilon^{u,V}_{ee}$, $\epsilon^{d,V}_{ee}$, and $\epsilon^{u,V}_{e\mu}$ nonzero, Eq.~\ref{eq:cevns_flav_degen} becomes the equation for an infinite elliptic cylinder in three dimensions. If we add in $\epsilon^{d,V}_{e\mu}$, cancellations through the linear combination $\epsilon^{u,V}_{e\mu} + \frac{2N+Z}{2Z+N}\epsilon^{d,V}_{e\mu}$ become available, once again yielding sets of plane solutions;
\begin{equation}
\label{eq:cevns_flav_degen2}
\left\{
    \begin{aligned}
        \epsilon^{u,V}_{ee} + \dfrac{2N+Z}{2Z+N}\epsilon^{d,V}_{ee} - \dfrac{Q_V}{2Z+N} &= 0 \quad\\
        \epsilon^{u,V}_{ee} + \dfrac{2N+Z}{2Z+N}\epsilon^{d,V}_{ee} &= 0 \\
        \epsilon^{u,V}_{e\mu} + \dfrac{2N+Z}{2Z+N}\epsilon^{d,V}_{e\mu} &= 0
    \end{aligned}
\right\}
\end{equation}

\subsubsection{E$\nu$ES Experiments}
Now consider the E$\nu$ES cross-section but with two NSI flavor indices nonzero, each with left and right chiral components ($\epsilon^{e,L}_{ee}$, $\epsilon^{e,R}_{ee}$, $\epsilon^{e,L}_{e\mu}$, and $\epsilon^{e,R}_{e\mu}$, for example). Since the current generation of experiments sensitive to E$\nu$ES are flavor-blind (just like in the CE$\nu$NS case), $e^- + \nu_e \rightarrow e^- + \nu_e$ and $e^- + \nu_e \rightarrow e^- + \nu_\mu$ have indistinguishable final states, in which case we have to sum the cross-section over final states. This gives rise to the following system of equations;
\begin{equation}
\left\{
\begin{aligned}
(1 + g_L + \epsilon^{e,L}_{ee})^2 + (g_R + \epsilon^{e,R}_{ee})^2 + (\epsilon^{e,L}_{e\mu})^2 + (\epsilon^{e,R}_{e\mu})^2 &= (1+g_L)^2 + g_R^2 \\
2(g_R + \epsilon^{e,R}_{ee})^2 + 2(\epsilon^{e,R}_{e\mu})^2 + \dfrac{m_e}{E_\nu}\bigg((g_R + \epsilon^{e,R}_{ee})(g_L + \epsilon^{e,L}_{ee}) + \epsilon^{e,R}_{e\mu}\epsilon^{e,L}_{e\mu}\bigg) &= 2g_R^2 + \dfrac{m_e}{E_\nu}g_R g_L \\
(g_R + \epsilon^{e,R}_{ee})^2 + (\epsilon^{e,R}_{e\mu})^2 &= g_R^2
\end{aligned}
\right\}
\end{equation}
One may check that this system of equations has a single real solution $\epsilon^{e,L}_{ee} = \epsilon^{e,R}_{ee} = \epsilon^{e,L}_{e\mu} = \epsilon^{e,R}_{e\mu} = 0$, provided $m_e / E_\nu$ remains of order 1. As $m_e / E_\nu$ tends to 0, the sets of curves defined by the above three equations no longer intersect in NSI space, which translates to a less resolved profile likelihood if one would perform a maximum likelihood estimation on neutrino scattering data sensitive to the E$\nu$ES cross-section.

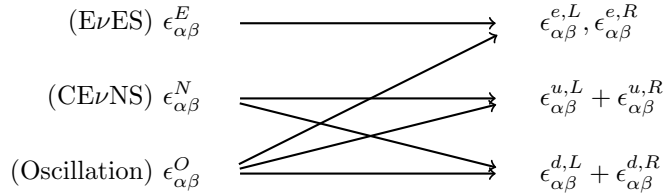
\begin{figure}[h]
 \centering
 \begin{tikzpicture}[ele/.style={fill=black,circle,minimum width=.8pt,inner sep=1pt},every fit/.style={ellipse,draw,inner sep=-2pt}]
  \node[label=left:(E$\nu$ES) $\epsilon^{E}_{\alpha\beta}$] (a1) at (0,4) {};    
  \node[label=left:(CE$\nu$NS) $\epsilon^{N}_{\alpha\beta}$] (a2) at (0,3) {};    
  \node[label=left:(Oscillation) $\epsilon^{O}_{\alpha\beta}$] (a3) at (0,2) {};

  \node[label=right:{$\epsilon^{e,L}_{\alpha\beta},\epsilon^{e,R}_{\alpha\beta}$}] (b1) at (4,4) {};
  \node[label=right:$\epsilon^{u,L}_{\alpha\beta}+\epsilon^{u,R}_{\alpha\beta}$] (b2) at (4,3) {};
  \node[label=right:$\epsilon^{d,L}_{\alpha\beta}+\epsilon^{d,R}_{\alpha\beta}$] (b3) at (4,2) {};

  \draw[->,thick,shorten <=4pt,shorten >=6] (a1) -- (b1);
  \draw[->,thick,shorten <=4pt,shorten >=6] (a2) -- (b2);
\draw[->,thick,shorten <=4pt,shorten >=6] (a2) -- (b3);
\draw[->,thick,shorten <=4pt,shorten >=6] (a3) -- (b1);
\draw[->,thick,shorten <=4pt,shorten >=6] (a3) -- (b2);
\draw[->,thick,shorten <=4pt,shorten >=6] (a3) -- (b3);
 \end{tikzpicture}
\caption{\label{fig:map} A cartoon of the dependencies of the phenomenological NSI on the physical NSI. E$\nu$ES experiments may gain sensitivity to the left and right chiral $e$-NSI, while CE$\nu$NS and oscillation experiments are only sensitive to the vectorial combinations of $u,d$ and $e,u,d$ NSI, respectively.}
\end{figure}

What we hope to convey at this stage is that as one continues to introduce more NSI parameters into the scattering cross-sections and oscillation Hamiltonian, the number of solutions may increase and become less resolved in a likelihood analysis, but the relationships between NSI of different fermion indices remains the same. In the spirit of Figure~\ref{fig:fermion_planes}, the neutrino oscillation matter potential, the CE$\nu$NS cross section, and the E$\nu$ES cross section form a weak mapping between the \textit{phenomenological} NSI $\epsilon$ to \textit{physical} NSI $\epsilon$ that exploited by combining the results of the three classes of experiments (Figure~\ref{fig:map}). In order to concretely establish sensitivity to the NSI operators that we have considered, it is essential to combine all three of these classes of experiments together in a global analysis.

\section{Combining Oscillation, CE$\nu$NS, and E$\nu$ES Data}\label{sec:priorflow}
Motivated by the degeneracy structures we have just discussed, we will now attempt to illustrate how CE$\nu$NS, E$\nu$ES, and oscillation experiments can be joined together in a global analysis. We will work under the pretense that NSI of all fermion indices $f=e,u,d$ are free to vary; in other words, none of them will be fixed to zero or other values on the basis of external limits. Doing this permits 24 NSI parameters in the count; 12 from $ \epsilon^{e,L}_{\alpha\beta}$ and $\epsilon^{e,R}_{\alpha\beta}$, 6 from $\epsilon^{u,V}_{\alpha\beta}$ and 6 from $\epsilon^{d,V}_{\alpha\beta}$. It is  reminded  that we have limited these NSI to be real-valued, and the inclusion of complex phases will be reserved for a later study which also involves the vacuum parameters such as the PMNS phase $\delta_{CP}$. As we outlined in Section~\ref{sec:intro}, we take this relatively large set of parameters for two key reasons that we will summarize again. The first reason is that one should not be limited by the pretense that if there is new physics in the neutrino sector, the new vector operators should be restricted to simple combinations of lepton non-universal or flavor-changing neutral currents. The second reason is primarily technological; that we would like to be able to include many NSI parameters (and many experimental data) to understand the full space of degeneracies with computational impunity.

To build our ensemble of experimental data for the global fit, we first begin by considering a minimal setup consisting of a single CE$\nu$NS dataset and a single E$\nu$ES dataset. The solar neutrino spectrum measured by Borexino in Phase II of the experimental program~\cite{Agostini:2018uly} gives us control over the $e$-NSI through E$\nu$ES, while the COHERENT collaboration's open data release and observation of the CE$\nu$NS interaction at a CsI detector~\cite{Akimov:2017ade} provides sensitivity to the $u$ and $d$ quark NSI. Analyses have been performed with these data sets in both minimal NSI scenarios and for a broad range of operators~\cite{AristizabalSierra:2018eqm, Khan:2019jvr,Giunti:2019xpr,Khan:2017oxw,Coloma:2019mbs,Agarwalla:2019smc,Denton:2018xmq, Miranda:2020zji}, but to explore the full space of NSI degeneracies we will allow an NSI set of maximum size to be free to vary, which has not been done before. Secondly, the stopped-pion spallation neutrino source (SNS) at COHERENT only produces $\nu_e$, $\nu_\mu$, and $\bar{\nu}_\mu$ neutrinos, and with a short baseline to the detectors, there is negligible oscillation of the neutrino flux into $\tau$ flavors. This implies a lack of sensitivity to $\epsilon^{u,V}_{\tau\tau}$ and $\epsilon^{d,V}_{\tau\tau}$ parameters.

To supplement our analysis with a dataset sensitive to $\epsilon^{u,V}_{\tau\tau}$ and $\epsilon^{d,V}_{\tau\tau}$ NSI, we extend the experiments considered so far to include future projections at a future liquid xenon (LXe) dark matter detector (DMD); being optimized for detecting nuclear interactions with the DM halo, a kiloton-scale detector would be sensitive enough explore NSI through the CE$\nu$NS and E$\nu$ES cross-sections from naturally occurring neutrino fluxes~\cite{Link:2019pbm,Dutta:2017nht,AristizabalSierra:2017joc,Baudis:2013qla}. This is the ``neutrino floor" that DM direct detection experiments are soon set to encounter, and may be enhanced in the presence of NSI~\cite{Boehm:2018sux}. We will project sensitivity to $u$ and $d$ NSI from atmospheric neutrinos, which have the full flavor-range to access $\epsilon^{u,V}_{\tau\tau}$ and $\epsilon^{d,V}_{\tau\tau}$ NSI parameters through enhancements to the CE$\nu$NS cross-section. Such an experiment would also be sensitive to solar neutrinos, which lie at energy ranges such that they mainly interact through neutrino-electron scattering, provided that they can discriminate between electron and nuclear recoils. Therefore we also project the sensitivity to $\epsilon^{e,L}_{\alpha\beta}$ and $\epsilon^{e,R}_{\alpha\beta}$ NSI with solar neutrinos at an LXe DMD to complement the Borexino analysis.

Finally, for the neutrino oscillations aspect of our analysis strategy, we will project simulated data at DUNE from atmospheric neutrinos oscillating through the Earth, which have been studied before in a variety of contexts for measuring the PMNS matrix and NSI~\cite{Coloma:2015kiu, Farzan:2017xzy}. We stress here that atmospheric neutrinos can supply DUNE with very rich oscillation data below $E_\nu = 1$ GeV~\cite{Kelly:2019itm} to supplement data from beam neutrinos. The atmospheric neutrino flux contains a host of all $e$, $\mu$, and $\tau$ flavor neutrinos after oscillation through the Earth's mantle, and effectively comprises a range of oscillation baselines as neutrinos propagate through the range of zenith angles. In addition, although it is not considered in this work, the hierarchy degeneracies and the interplay of the CP phase $\delta_{CP}$ in the PMNS matrix with NSI can be explored in DUNE~\cite{Deepthi:2016erc}.

\subsection{Prior-flow}
Now we will outline the statistical treatment for the simulated and measured data in the experiments considered. We take a Bayesian inference approach to constraining the NSI parameters. In the following discussion we use the Bayesian inference package \texttt{MultiNest}~\cite{Feroz2009} to construct likelihood functions and compute the posterior probability distributions of our NSI parameters given the simulated data at each experiment.

To combine the likelihood information from several experiments, traditionally what is done is the construction of a single global likelihood function calculated by the simultaneous simulation of the event spectra for each class of experiment considered. In the context of this analysis, the joint likelihood function would take physical NSI as model inputs, giving $\mathcal{L} = P(\mathcal{D} | \vec{\epsilon}, H)$ for NSI parameters $\vec{\epsilon}$, observed data $\mathcal{D} = \cup_{i=1}^N \mathcal{D}_i$ for experiments $i=1,\dots,N$, and a null hypothesis $H$ for which we take all NSI as zero, i.e. $\vec{\epsilon} = \vec{0}$. This style of approach has been taken before in a variety of global analysis settings for neutrino NSI~\cite{Khan:2016uon, Esteban:2018ppq, Esteban:2019lfo}. This approach can be very computationally expensive, and with 24 parameters in our consideration (12 from $ \epsilon^{e,L}_{\alpha\beta}$ and $\epsilon^{e,R}_{\alpha\beta}$, 6 from $ \epsilon^{u,V}_{\alpha\beta}$, and 6 from $ \epsilon^{u,V}_{\alpha\beta}$) it may be difficult to accurately discover the posterior distribution in such a large prior volume, let alone converge at all in the evidence computation. This is even without taking into account the potentially many experimental nuisance parameters for detector response, background or signal uncertainties, etc., in the likelihood fit. In fact, in a typical global analysis we need not allow so many NSI parameters to be nonzero at the same time; experimental nuisance parameters can be enough to create a likelihood parameter space of relatively large dimension.

Instead, we take a divide-and-conquer approach illustrated as follows. Suppose we aim to measure NSI parameters $x$ and $y$. Then suppose we posses or simulate data from two experiments A and B such that experiment A is sensitive to $x$ and $y$ while experiment B is sensitive to $y$ alone. One can then use experiment B to measure $y$ and its posterior distribution given the data at B, $\pi(y\mid \mathcal{D}_B)$, and subsequently take this posterior distribution as the prior distribution on $y$ for experiment A. In the context of Bayes theorem,
\begin{equation}
    \pi(x,y \mid \mathcal{D}_A) = \dfrac{\mathcal{L}(\mathcal{D}_A \mid x,y ; H) \cdot \{\pi(y\mid \mathcal{D}_B) \cdot u(x)\}}{Z},
\end{equation}
where $Z$ is the Bayesian evidence. Since experiment B provides no information on $x$, we take a uniform prior density $u(x)$ over an appropriate interval such that the joint prior becomes $$\pi(x,y)=\pi(y\mid\mathcal{D}_B)\cdot u(x).$$ By doing this, we effectively constrain $y$ at experiment A using its prior distribution from experiment B. If A and B have complementarity between any degeneracies that may exist for $x$ and $y$, they would be combated just as they would by directly calculating $\pi(x,y\mid \mathcal{D}_A\cup \mathcal{D}_B)$ via a joint likelihood function for the data from the two experiments A and B.
This strategy can be repeated for numerous experiments and with more parameters. By allowing posterior information to ``flow" from one experiment into another, we effectively reduce the prior volume that needs to be searched. This is what we propose using data from the COHERENT and Borexino collaborations, in addition to projections for a future LXe DMD and the DUNE experiment, bearing in mind that this scheme could be extended to a variety of others.

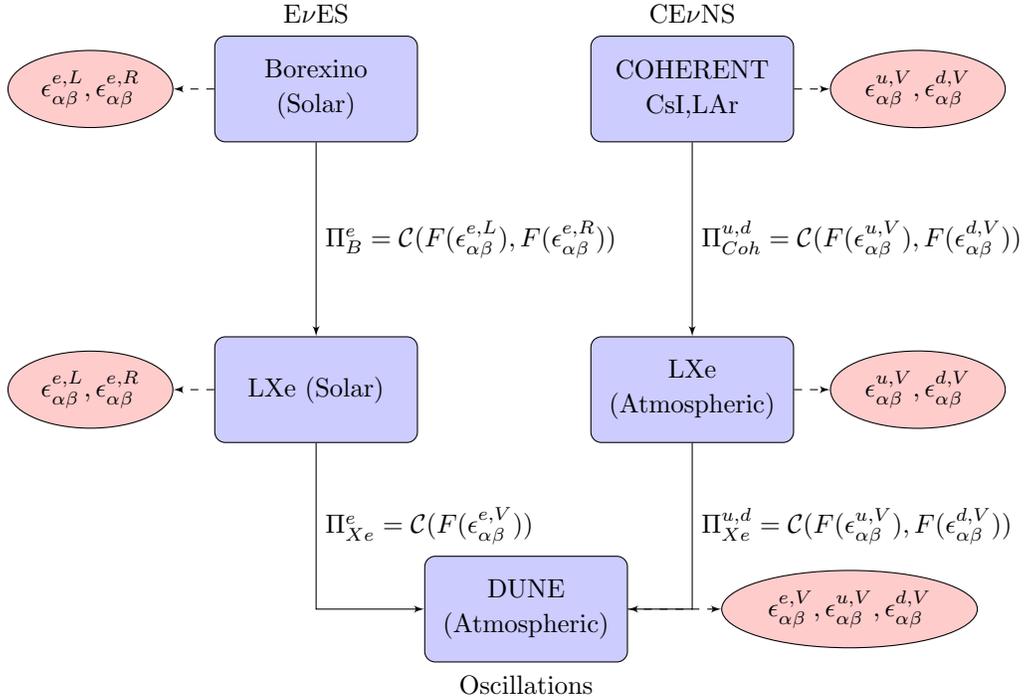
\begin{figure*}
    \centering
\begin{tikzpicture}[node distance = 4cm, auto]
    \node [block] (coh) {COHERENT\\CsI,LAr};
    \node [block, below of=coh] (xen) {LXe \\(Atmospheric)};
    \node [block, left of=coh, node distance=5cm] (borex) {Borexino (Solar)};
    \node [block, left of=xen, node distance=5cm] (xee) {LXe (Solar)};
    \node [block, below left=1.5cm and -0.5cm of xen] (dune) {DUNE \\(Atmospheric)};
    \node [cloud, left of=borex, node distance=3cm] (borex_label) {$\epsilon^{e,L}_{\alpha\beta},\epsilon^{e,R}_{\alpha\beta}$};
    \node [cloud, left of=xee, node distance=3cm] (xee_label) {$\epsilon^{e,L}_{\alpha\beta},\epsilon^{e,R}_{\alpha\beta}$};
    \node [cloud, right of=xen, node distance=3cm] (xen_label) {$\epsilon^{u,V}_{\alpha\beta}, \epsilon^{d,V}_{\alpha\beta}$};
    \node [cloud, right of=dune, node distance=4.3cm] (dune_label) {$\epsilon^{e,V}_{\alpha\beta},\epsilon^{u,V}_{\alpha\beta}, \epsilon^{d,V}_{\alpha\beta}$};
    \node [cloud, right of=coh, node distance=3cm] (coh_label) {$\epsilon^{u,V}_{\alpha\beta}, \epsilon^{d,V}_{\alpha\beta}$};
    \node [above of=coh, node distance=1cm] (cevns) {CE$\nu$NS};
    \node [above of=borex, node distance=1cm] (vees) {E$\nu$ES};
    \node [below of=dune, node distance=1cm] (osc) {Oscillations};
    \path [line] (coh) -- node {$\Pi^{u,d}_{Coh} = \mathcal{C}(F(\epsilon^{u,V}_{\alpha\beta}),F(\epsilon^{d,V}_{\alpha\beta}))$} (xen);
    \path [line] (xen) |- node [near start] {$\Pi^{u,d}_{Xe} = \mathcal{C}(F(\epsilon^{u,V}_{\alpha\beta}),F(\epsilon^{d,V}_{\alpha\beta}))$} (dune);
    \path [line] (borex) -- node {$\Pi^{e}_B = \mathcal{C}(F(\epsilon^{e,L}_{\alpha\beta}),F(\epsilon^{e,R}_{\alpha\beta}))$} (xee);
    \path [line] (xee) |- node [near start] {$\Pi^{e}_{Xe} = \mathcal{C}(F(\epsilon^{e,V}_{\alpha\beta}))$} (dune);
    \path [line, dashed] (dune) -- (dune_label);
    \path [line, dashed] (xee) -- (xee_label);
    \path [line, dashed] (borex) -- (borex_label);
    \path [line, dashed] (xen) -- (xen_label);
    \path [line, dashed] (coh) -- (coh_label);
\end{tikzpicture}
\caption{The ``Prior-Flow" of joint probability information from experiment to experiment. We begin with E$\nu$ES from solar neutrinos at Borexino and CE$\nu$NS at COHERENT, then proceed to E$\nu$ES and CE$\nu$ES scattering at a future 1 kton$\cdot$year LXe dark matter detector, and finally to future atmospheric neutrino oscillation measurements at DUNE. The components of each prior that are inherited from a previous experiment are denoted by $\Pi$, which are taken as empirical copulas of the relevant marginals of the NSI parameters. If one of the previous experiments is not sensitive to a particular NSI, the uniform distribution is taken as a prior for that corresponding parameter.}
\label{fig:priorflow}
\end{figure*}

The prior ordering structure is shown in Figure~\ref{fig:priorflow}. At the top level, only uniform priors are used, and by default we fix the uniform interval to $(-1, 1)$ for all vector NSI (and $(-0.5, 0.5)$ for $L$ and $R$ components) for the sake of simplicity. Each subsequent experiment in the ``prior-flow" takes its prior from the joint posterior distributions of the experiments above, for the relevant subset of NSI to which those experiments are sensitive, with uniform priors for the NSI that remain. The explicit sets of $\vec{\epsilon}$ and their priors are listed in Table~\ref{tab:nsi}.
\begin{table*}[t]
  \centering
  \caption{\label{tab:nsi} NSI parameters used in this analysis and the prior scheme used for each experiment. Here $U_N$ are $N$-dimensional uniform priors on the NSI vector of length $N$, chosen to range from $-0.5$ to $0.5$. $\Pi^e_B$, $\Pi^{u,d}_C$, and $\Pi^{u,d}_{Xe}$ are the priors taken from the posterior distributions at Borexino, COHERENT, and a future LXe DMD, respectively.}
  \begin{tabular}{ l l l c }
 Experiment & NSI & Flavor Indices $(\alpha,\beta)$ & Prior \\
 \hline
  Borexino (solar) & $\epsilon^{e,L}_{\alpha\beta} , \epsilon^{e,R}_{\alpha\beta}$ & $ee,\mu\mu,\tau\tau,e\mu,e\tau,\mu\tau$  &$U_{12}$ \nonumber \\
      LXe (solar) & $\epsilon^{e,L}_{\alpha\beta} , \epsilon^{e,R}_{\alpha\beta}$ & $ee,\mu\mu,\tau\tau,e\mu,e\tau,\mu\tau$  &$\Pi^{e}_{B}$ \nonumber \\
   COHERENT (LAr / CsI) & $\epsilon^{u,V}_{\alpha\beta} , \epsilon^{d,V}_{\alpha\beta}$ & $ee,\mu\mu,e\mu,e\tau,\mu\tau$  &$U_{10}$ \nonumber \\
   LXe (atmos.) & $\epsilon^{u,V}_{\alpha\beta} , \epsilon^{d,V}_{\alpha\beta}$ & $ee,\mu\mu,\tau\tau,e\mu,e\tau,\mu\tau$  &$\Pi^{u,d}_{C} \otimes U^{u}_{\tau\tau}\otimes U^{d}_{\tau\tau}$ \nonumber \\
     DUNE (atmos.) & $\epsilon^{e,V}_{\alpha\beta}, \epsilon^{u,V}_{\alpha\beta} , \epsilon^{d,V}_{\alpha\beta}$ & $ee,\mu\mu,\tau\tau,e\mu,e\tau,\mu\tau$  & $\Pi^{u,d}_{Xe} \otimes \Pi^{e}_{Xe}$ \nonumber
  \end{tabular}
\end{table*}

\subsection{Copulas}
With such a strategy, there is an important question as to how one models the prior distributions of a multivariate set of NSI parameters. We remind the reader that according to our strategy, this joint prior of the parameters is constructed  based on the posterior distributions of the previous experiments.

If one only uses the one-dimensional marginal distribution as individual prior on each parameter, important correlations between the NSI will be lost. Therefore, we elect to model the joint prior distribution as completely as possible. To do this, we use a copula. In $d$ dimensions, a copula $\mathcal{C}$ is a cumulative distribution function (CDF) $\mathcal{C}:[0,1]^d \to [0,1]$ with uniform marginal distributions. See~\cite{nelsen2006,Sklar1973RandomVJ} for a review. Sklar's theorem \cite{Sklar1959} states that for every $d$-dimensional joint CDF, in our case $\mathcal{F}(\epsilon_1,\dots,\epsilon_d)$ for NSI parameters $\epsilon_1,\dots,\epsilon_d$, there exists a $d$-copula $\mathcal{C}$ such that
\begin{equation}
    \mathcal{F}(\epsilon_1,\dots,\epsilon_d) = \mathcal{C}(F_1(\epsilon_1),\dots,F_d(\epsilon_d))
\end{equation}
where $F_1,\dots,F_d$ are the marginal distributions of the NSI parameters. 
Copula functions, in essence, connect the marginal distributions and the joint distribution through a correlation structure. Given absolutely continuous marginal distributions and the joint distribution, the copula function is unique.

The copula $\mathcal{C}$ is usually a function, which can sometimes be written in closed form, whose form is associated with the dependency structures of a known family of statistical distributions. There are many families of copula, and no single copula is guaranteed to be a perfect model of the underlying joint distribution, so in practice one usually chooses the family that best fits the sample data of the joint distribution. For example, the band-shaped degeneracy contours between pairs of $\epsilon^{u,V}_{\alpha\beta}$ and $\epsilon^{d,V}_{\alpha\beta}$ NSI (which we will see in Section~\ref{sec:results}) may be well-modeled by the Frank family of copulas that captures this kind of correlation well.
However, one may also use an \textit{empirical} copula to fit the joint prior distribution provided one has sample data from \texttt{MultiNest}. We elect to use this option to fit the posterior joint distributions and to subsequently simulate prior distributions in the prior-flow, since this is the most robust and accurate way of modeling the NSI dependency structure discovered by each experiment.
To do this we use the \texttt{R} package \texttt{copula} to fit empirical copulas to the joint distributions for each experiment in the prior flow. 

\begin{figure}[h]
  \begin{center}
   \includegraphics[width=0.32\textwidth]{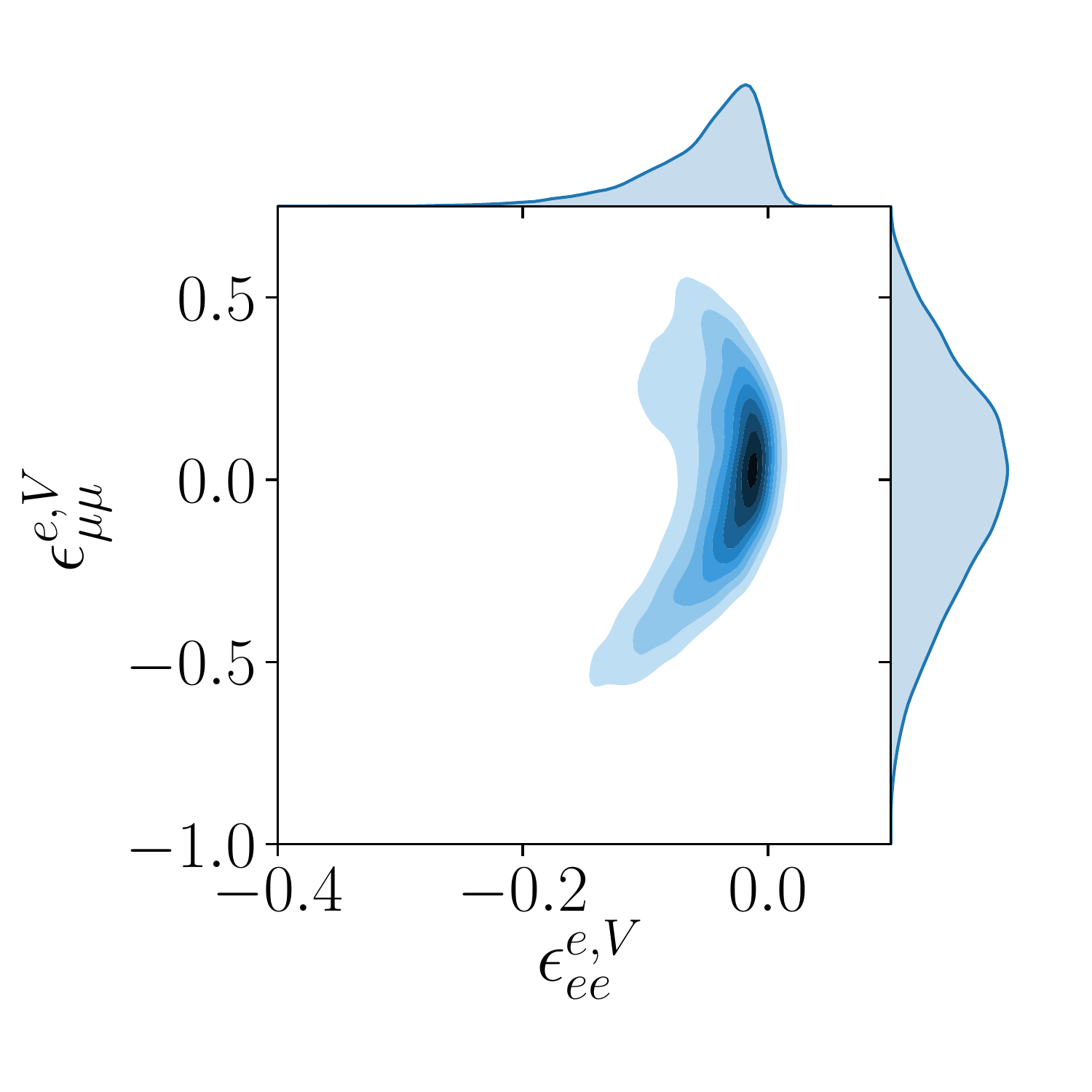}
   \includegraphics[width=0.32\textwidth]{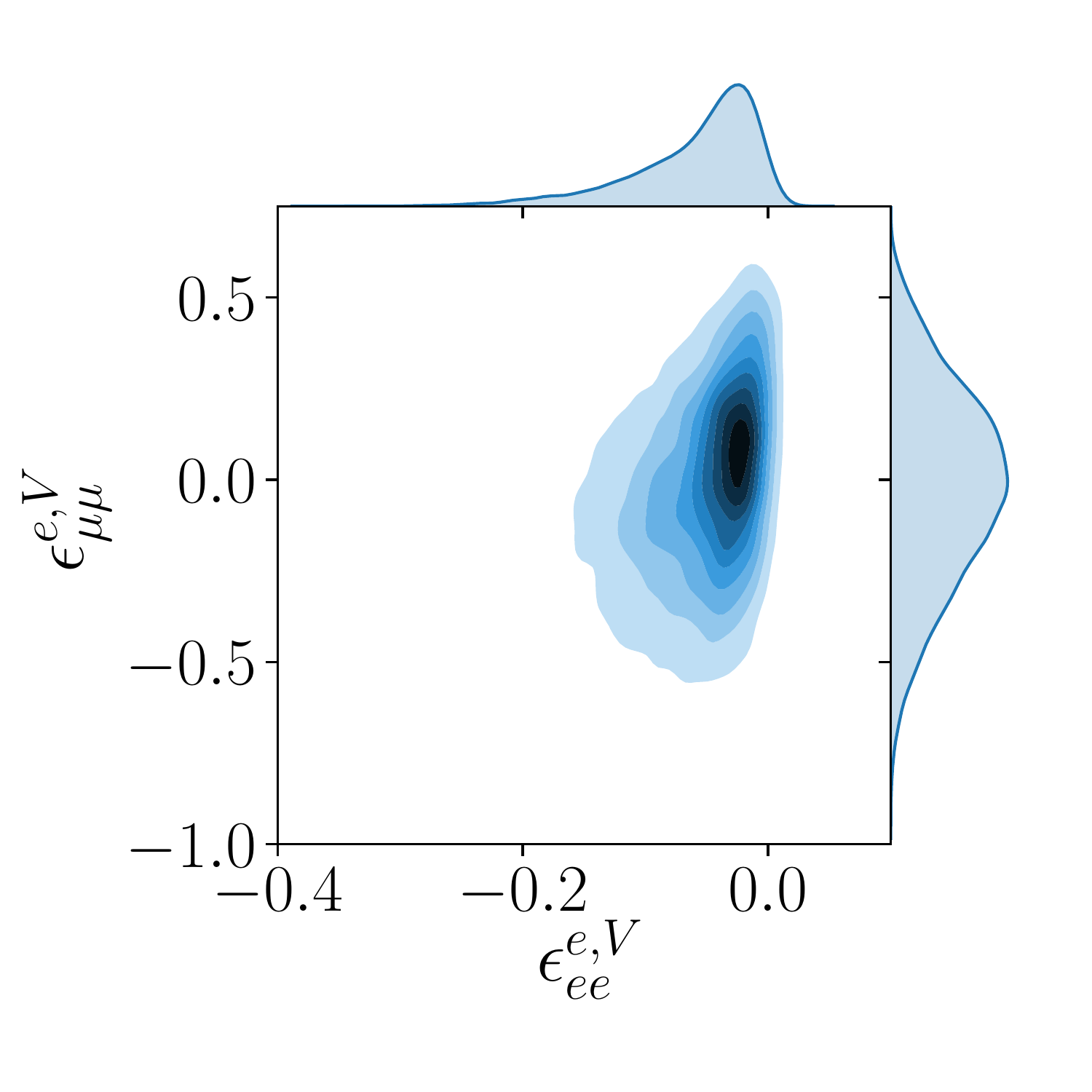}\\
   \includegraphics[width=0.32\textwidth]{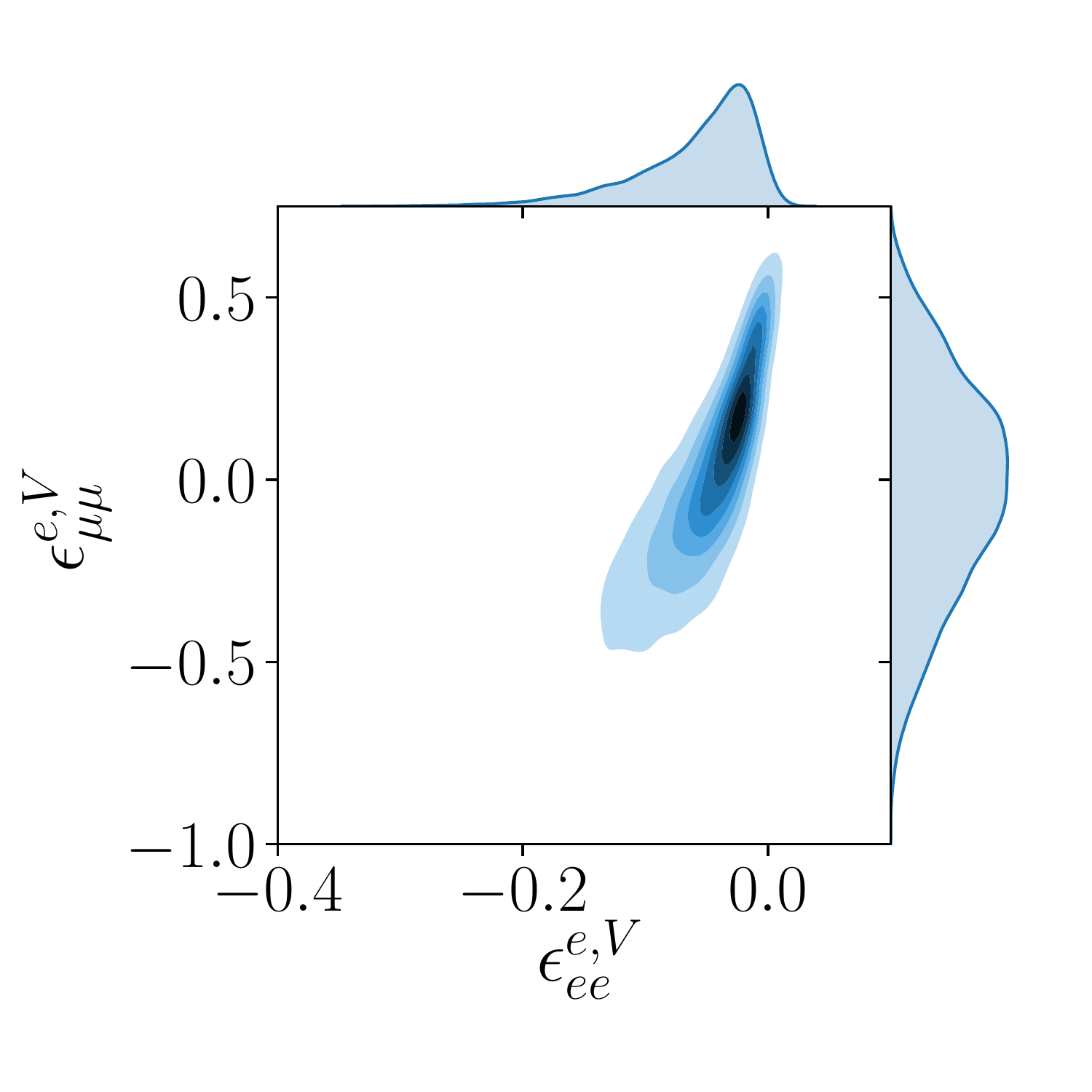}
   \includegraphics[width=0.32\textwidth]{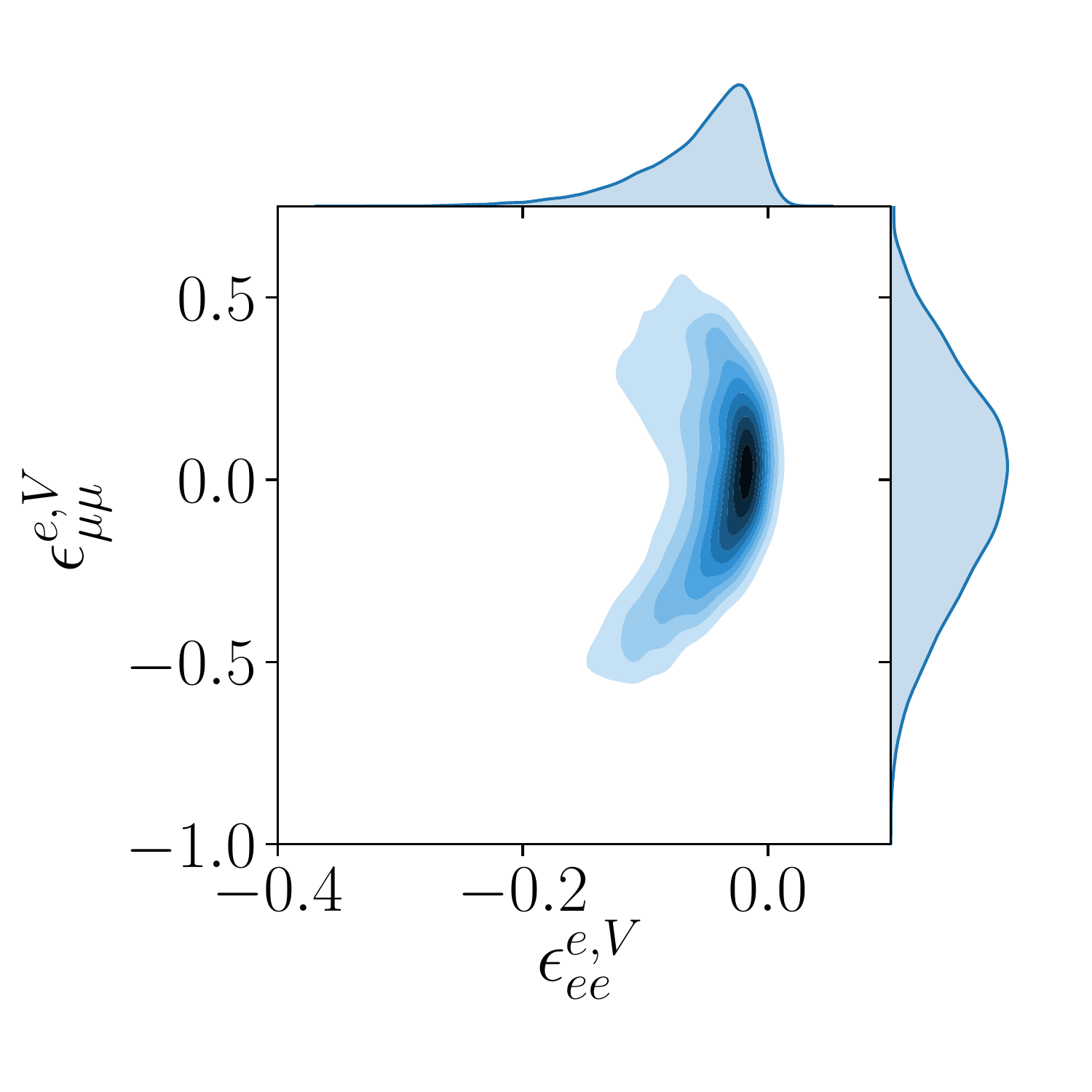}
    \caption{\label{fig:multinest_vs_copula} Posterior kernel density contours and their marginals plotted for the $\epsilon^{e,V}_{ee}$-$\epsilon^{e,V}_{\mu\mu}$ NSI. The joint distribution observations from \texttt{MultiNest} (top-left) are compared with $10^{4}$ simulated samples drawn from fits for a Gaussian copula (top-right), a Gumbel copula (bottom-left), and an empirical copula (bottom-right). In each fit, the 1-D marginals are reproduced well, but only the empirical copula can accurately reproduce the non-trivial dependency structure of the \texttt{MultiNest} observations in this example.}
    \end{center}
\end{figure}

In \texttt{MultiNest}, the prior is formally implemented as a map from the $d$-dimensional cube $[0,1]^d \to \mathbf{R}^d$ via the inverse CDF or quantile function of the prior distribution. This naturally lends itself to the implementation of copulas, since the available methods of simulation (finding the inverse of a multivariate CDF) are well documented. In order to simulate samples from the empirical copulas, we employ the \textit{conditional distribution method}~\cite{nelsen2006,johnson1987} followed by using the inverse CDFs of the prior marginals to extract sample NSI parameters for each iteration in \texttt{MultiNest}. For an empirical copula $\mathcal{C}$ this procedure is illustrated as follows;
\begin{align}
    1. & \text{ \texttt{MultiNest} generates } u_1,\dots,u_d \sim U(0,1) \nonumber \\
    2. & u_2 \to \mathcal{C}_{2\mid1}^{-1}(u_2 \mid u_1) \nonumber \\
        & u_3 \to \mathcal{C}_{3\mid2,1}^{-1}(u_3 \mid u_1, u_2) \nonumber \\
        & \dots \nonumber \\
        & u_d \to \mathcal{C}_{d\mid d-1,\dots,1}^{-1}(u_d \mid u_1,\dots,u_{d-1}) \nonumber \\
    3. & \text{ Set } x_1 = F_1^{-1} (u_1) ,\dots, x_d = F_d^{-1} (u_d) \nonumber
\end{align}
where the $\{x_i\}$, $i=1,\dots,d$ correspond to the $d$ NSI parameters used in each prior and their 1-dimensional marginal CDFs $F_1,\dots,F_d$. In step 2 we compute the the conditional distributions of the copula, $\mathcal{C}_{r\mid r-1,\dots,1}$, which is equivalent to finding its partial derivatives and taking the pseudo-inverse~\cite{Strelen2007AnalysisAG}. Since the empirical copula $\mathcal{C}$ and its partial derivatives are numerically computed from the \texttt{MultiNest} observations, this inverse is found numerically as well. As an example of how closely the empirical copula can model one of the joint posterior distributions on the NSI, a comparison between \texttt{MultiNest} samples and the corresponding empirical copula simulation is shown in Figure~\ref{fig:multinest_vs_copula}.

\section{Methods}\label{sec:methods}
\subsection{Stopped-pion neutrinos at COHERENT}
There have been numerous analyses of COHERENT neutrino data for the determination of NSI, neutrino vacuum parameters and nuclear structure   \cite{AristizabalSierra:2019zmy,Denton:2018xmq,Cadeddu:2019eta,Cadeddu:2017etk, AristizabalSierra:2017joc, AristizabalSierra:2018eqm, Coloma:2019mbs, Giunti:2019xpr, Miranda:2020zji}. Particularly, we use  the procedure detailed in Ref.~\cite{Dutta:2019eml} which combines both energy and timing data in the neutrino spectrum at COHERENT from the 4466 kg$\cdot$days of exposure at the CsI detector. By using probability density functions (PDFs) of the time and energy spectra for the $\nu_\mu$, $\nu_e$, and $\bar{\nu_\mu}$ flavor components we can predict the number of observed prompt and delayed neutrino counts. To compute the number of events between recoil energies $E^a_r$ and $E^b_r$, we convolve the neutrino flux for each neutrino species $\alpha$ with the CE$\nu$NS cross-section and detector efficiency $\eta(E_r)$;
\begin{equation}
\label{eq:cevns_rates}
N = \dfrac{\mathcal{E}}{M_T} \sum_{\alpha} \int_{E_r^a}^{E_r^b} dE_r \int_{E_\nu^{min}}^\infty  \dfrac{d\Phi_{\nu_\alpha}}{dE_\nu} \dfrac{d\sigma}{dE_r} \eta(E_r) dE_\nu
\end{equation}
Here we take $E_\nu^{min} =  (\sqrt{E_r^2 + 2 m_N E_r}+ E_r)/2$ for a nucleus mass $m_N$ and neutrino fluxes $d\Phi_{\nu_\alpha}/dE_\nu$ for $\nu_\mu, \bar{\nu}_\mu$, and $\nu_e$ flavor neutrinos. 

\begin{figure}
    \centering
    \includegraphics[width=0.48\textwidth]{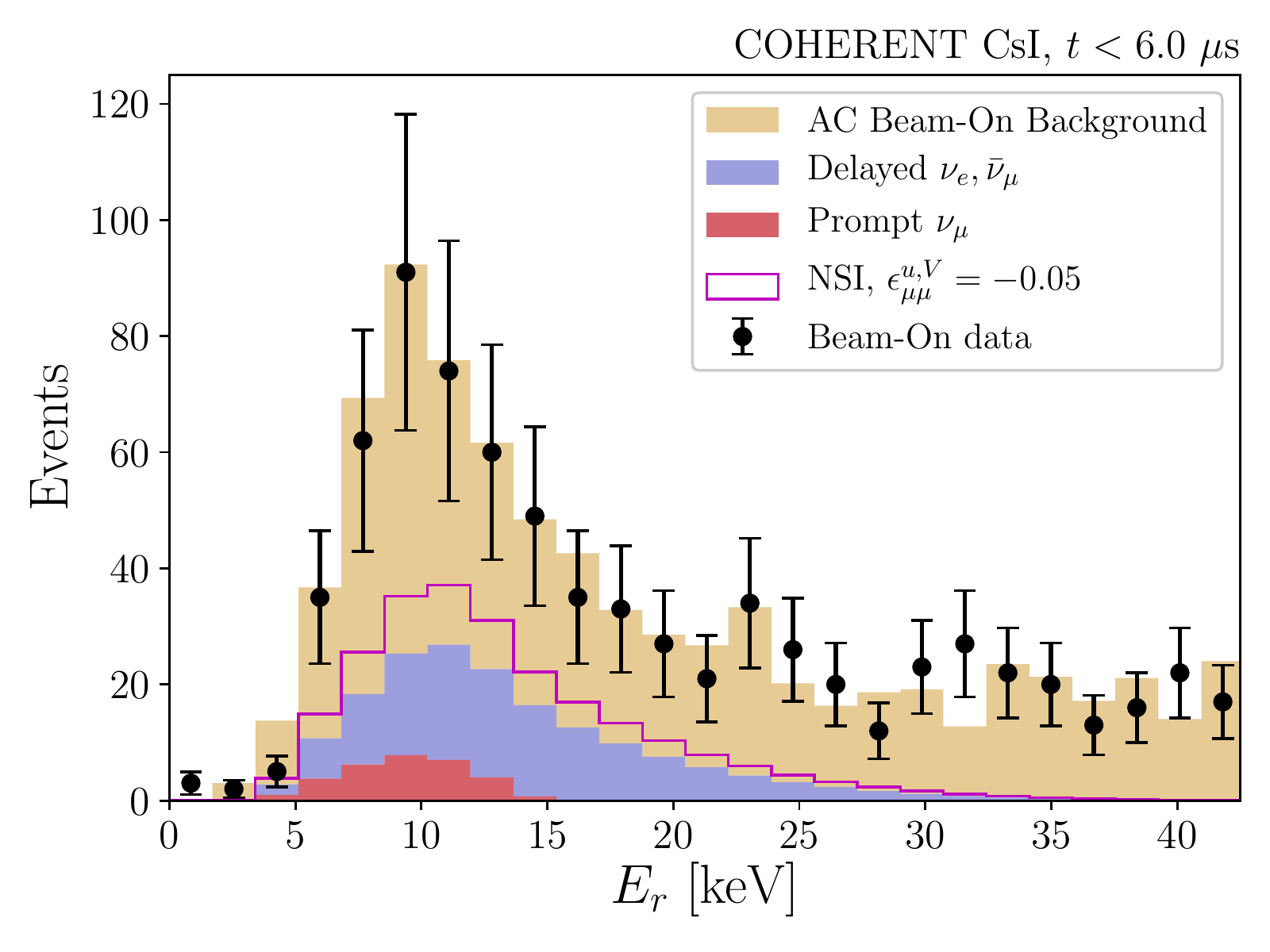}
    \includegraphics[width=0.48\textwidth]{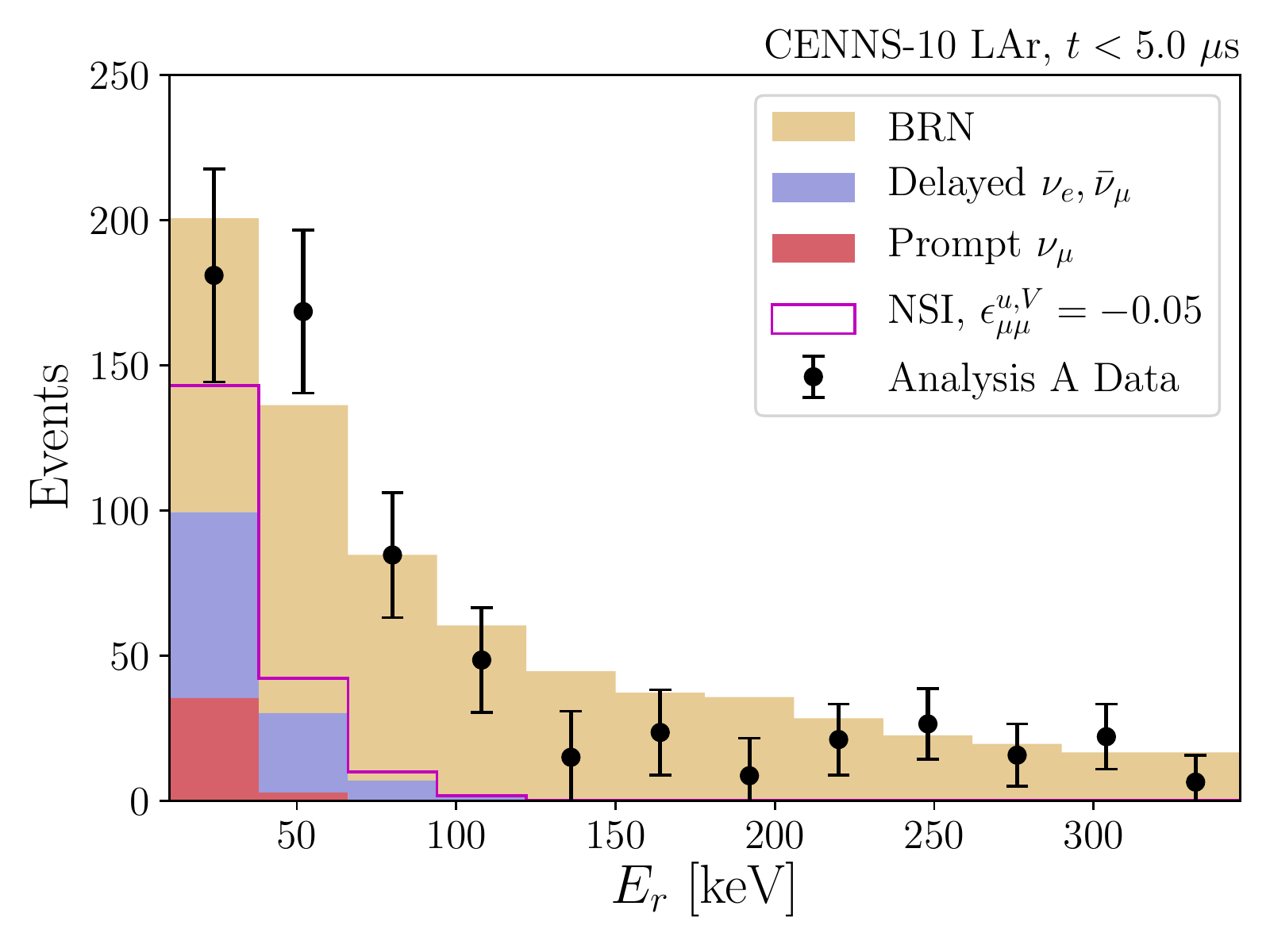}
    \caption{The COHERENT CsI event spectra (left) and the CENNS-10 LAr event spectra from ``Analysis A" (right) data is shown along with the standard neutrino interactions event rate prediction and BRN rates stacked together. Neutrino NSI event rates are separately shown for comparison.}
    \label{fig:coherent_spectra}
\end{figure}

To take advantage of the multiple detector materials available in by the SNS, in addition the data from the CsI detector we also include data from the liquid argon (LAr) CENNS-10 detector~\cite{Akimov:2020pdx}. Including an argon detector in the analysis will give the CsI CE$\nu$NS measurement some complementarity in the $(\epsilon^{u,V}_{\alpha\beta},\epsilon^{d,V}_{\alpha\beta})$ plane as discussed in Section~\ref{sec:fermion_degen}. Spectra for both the CsI and LAr analysis are shown in Figure~\ref{fig:coherent_spectra}.

In the CENNS-10 data release~\cite{Akimov:2020data}, data is binned not only in recoil energies $E_r$ and trigger times $t$, but also in a third dimension, $F_{90}$, corresponding to the light yield fraction in the first 90 ns of the PMT response. Therefore, we predict the 3D binned events in LAr for the using both timing and $F_{90}$ PDFs convoluted together with the energy response as in Eq.~\ref{eq:cevns_rates}.

For the CsI likelihood analysis, we again refer to Ref.~\cite{Dutta:2019eml} and use a binned log-likelihood that is marginalized over nuissance parameters for the backgrounds (steady-state (SS) and beam-related neutrons (BRN)) and systematic uncertainties.

For the LAr likelihood analysis, alternate PDFs are provided in the data release to encapsulate systematic uncertainties in the BRN and CE$\nu$NS rates. This allows us to add parameters to vary the expected CE$\nu$NS and BRN rates on top of the NSI variation as follows. First, there are several normalization uncertainties; $13\%$ uncertainty on the CE$\nu$NS normalization, $30\%$ and $100\%$ uncertainty on the prompt and delayed BRN components, respectively, and a modified statistical uncertainty of $\sqrt{N_{SS}/5}$ for the steady-state (SS) background (in accordance with measurements on SS done in a $\times 5$ oversampled time window). There are several systematics uncertainties; in the BRN timing mean and width, in the BRN energy distribution, in the CE$\nu$NS timing mean, and the CE$\nu$NS $F_{90}$-Energy distribution. For each of these, there is an alternate PDF template that represents the $\pm 1\sigma$ deviation from the BRN or SM CE$\nu$NS expectation. We represent the counts in these 3D PDFs over $E_r$, $F_{90}$, and $t$ as vector $\mathbf{n}$ over the flattened bins $i = 1,\dots,M$. Then, for each systematic uncertainty, we parameterize fluctuations in the expected PDFs as transformations of bin contents according to a normal (skew-normal) distribution given by the symmetric (asymmetric) systematic deviations in each bin~\footnote{By asymmetric, we mean that the variation of the systematic parameter by $\pm 1\sigma$ does not necessarily map onto equal excesses or deficits over the default PDF bin values. In practice, the CDF $F$ in Eq.~\ref{eq:coh_syst} should be checked such that after the bin-wise transformations, a $\pm1\sigma$ deviation in $u$ matches with the $\pm 1\sigma$ alternative PDFs.};
\begin{equation}
    n_i \to (1 + F^{-1} (u; \sigma_i, \alpha_i)) n_i,
    \label{eq:coh_syst}
\end{equation}
where $F^{-1}$ is the inverse-CDF of the normal wit mean 0 and standad deviation of $\sigma_i$ 
(or skew-normal distribution with the location parameter 0, a scale of  $\sigma_i$ and the skewness parameter  $\alpha_i$), whose argument $u\sim U(0,1)$ is a uniform variate allowed to float in the \texttt{MultiNest} likelihood scan. After the likelihood scan, the systematic and statistical nuisance parameters will be marginalized over and the NSI posterior distributions will be passed in as priors into the LXe fit as described in Figure~\ref{fig:priorflow}.

For the NSI parameters in both likelihoods, we take the real-valued $u$ and $d$ NSI as model inputs. We include $\epsilon^{u,V}_{ee}$, $\epsilon^{u,V}_{\mu\mu}$, $\epsilon^{u,V}_{e\mu}$, $\epsilon^{u,V}_{e\tau}$, $\epsilon^{u,V}_{\mu\tau}$, $\epsilon^{d,V}_{ee}$, $\epsilon^{d,V}_{\mu\mu}$, $\epsilon^{d,V}_{e\mu}$, $\epsilon^{d,V}_{e\tau}$, and $\epsilon^{u,V}_{\mu\tau}$, but not $\epsilon^{u,V}_{\tau\tau}$ or $\epsilon^{d,V}_{\tau\tau}$ because the negligible presence of $\tau$-flavor neutrinos at the SNS.

\subsection{Solar Neutrinos at Borexino}
The Borexino collaboration has measured the solar neutrino energy spectrum~\cite{Agostini:2018uly} over 92.1 kton$\cdot$days which provides an important dataset to help constrain NSI in neutrino-electron scattering events. We follow Refs.~\cite{Bellini:2013lnn,Agarwalla:2019smc} to model the solar neutrino energy spectrum at Borexino.
The solar neutrino event rate is predicted using the E$\nu$ES cross-section (Eq.~\ref{eq:eves}) and convolving it with the oscillated solar neutrino flux;
\begin{equation}
N = \sum_{\alpha,\beta,\gamma} \int_{E_r^a}^{E_r^b} \int_{E_\nu^{min}}^\infty \dfrac{\mathcal{E}}{M_T}  \dfrac{d\Phi_\alpha}{dE_\nu}  P_{\alpha\beta} \dfrac{d\sigma_{\beta\gamma}(E_r, E_\nu)}{dE_r} dE_r dE_\nu 
\label{eq:borex_rate}
\end{equation}
where we assume no direction reconstruction on the incoming neutrino and no flavor-sense, hence, the incoming, oscillated, and transition flavors $\alpha$ $\beta$ and $\gamma$ are summed over. The minimum neutrino energy is the same as the one used in Eq.~\ref{eq:cevns_rates} but with the replacement $m_N \to m_e$. 

We select data about the $^7$Be compton edge, corresponding to recoil energies from 550 keV to 1 MeV. As mentioned in Section~\ref{sec:fermion_degen}, this energy range is sensitive to NSI contributions to the $m_e / E_\nu$ proportional terms in the E$\nu$ES cross-section, which in turn helps converge on a single NSI solution during the likelihood analysis. This region also contains contributions from radiochemical backgrounds; $^{85}$Kr, $^{210}$Po, $^{11}$C, and $^{210}$Bi.

A log-likelihood function is constructed from the Borexino Phase II data $N^o_i$, while the  error standard deviation and the expected  number of events are denoted by $\sigma_i$ and $N^{s+b}_i$, respectively. 
The final likelihood combines information from all energy bins $i=1, 2, \dots$. We allow the predicted event rate to vary as a function of $\vec{k}=(k_{Be},k_{Po},k_{Kr},k_{Bi},k_{C})$ which parametrizes uncertainties in the $^7$Be flux and background rates. For a background rate $R_j$, we allow it to fluctuate via the parameter $k_j$ as follows;
\begin{equation}
    R_j \rightarrow (1+k_j) \cdot R_j
\end{equation}
We then take the Gaussian prior for these nuisance parameters $\vec{k}$ with means of 0 and widths given by the rate uncertainties. The predicted event rate is of course a function of NSI as well, taking $\vec{\epsilon} = (\epsilon^{e,L}_{ee},\epsilon^{e,L}_{ee},\epsilon^{e,L}_{\mu\mu},\epsilon^{e,L}_{\tau\tau},\epsilon^{e,L}_{e\mu},\epsilon^{e,L}_{e\tau},\epsilon^{e,R}_{\mu\tau},\epsilon^{e,R}_{\mu\mu},\epsilon^{e,R}_{\tau\tau},\epsilon^{e,R}_{e\mu},\epsilon^{e,R}_{e\tau},\epsilon^{e,R}_{\mu\tau})$. The log-likelihood function is now given in Eq.~\ref{eq:loglike}.
\begin{equation}
\label{eq:loglike}
    \ell = \sum_{i} \bigg\{-\dfrac{(N^{s+b}_i(\vec{k},\vec{\epsilon}) - N^o_i)^2}{2\sigma_i^2} - \dfrac{1}{2}\ln(2\pi\sigma_i)\bigg\}
\end{equation}

\begin{figure}[!tbh]
  \begin{center}
  \includegraphics[width=0.6\textwidth]{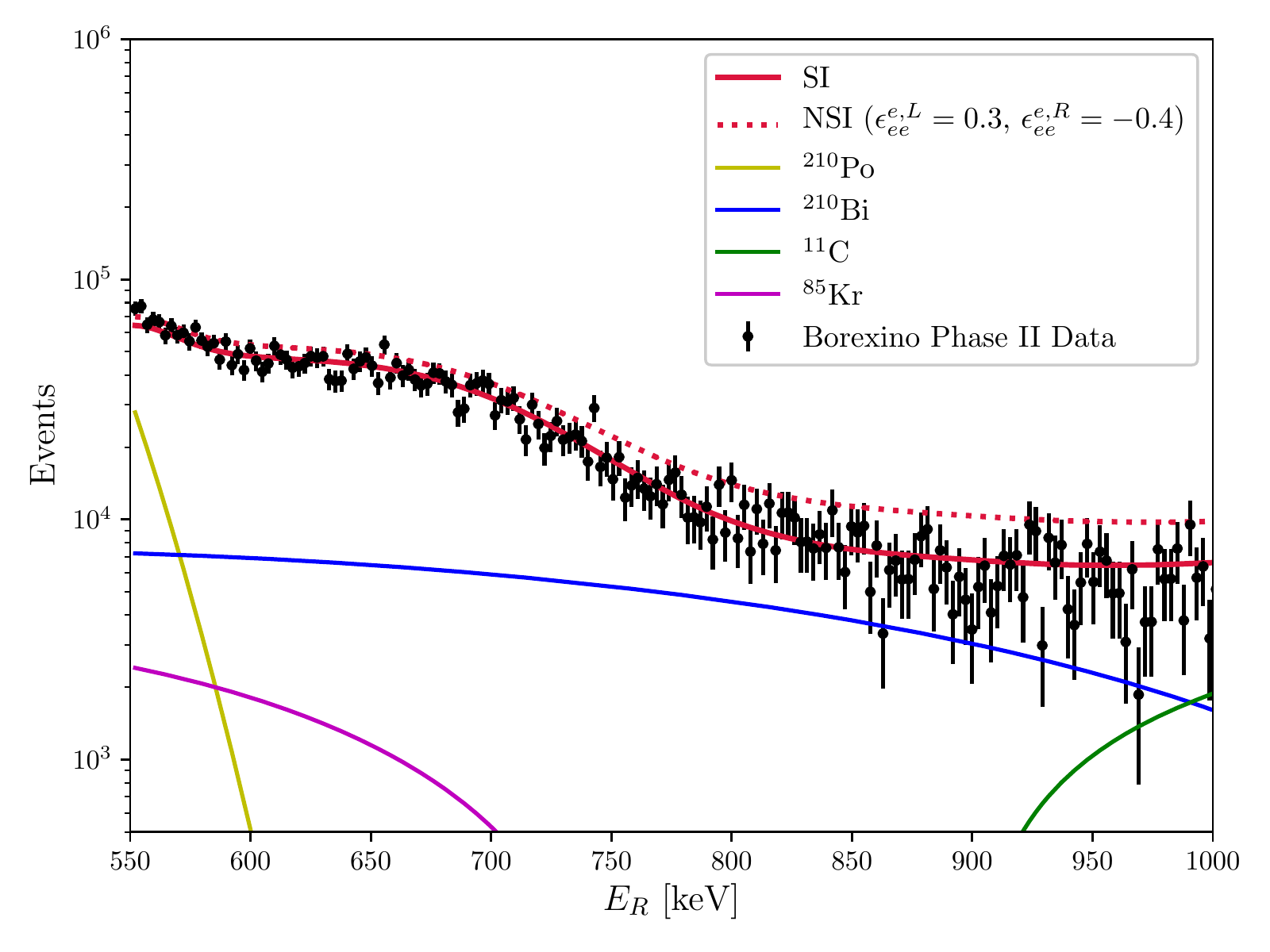}
    \caption{\label{fig:borex_solar} The solar neutrino spectrum from the Borexino Phase II dataset is shown around the $^7$Be compton edge. Data, and information on backgrounds, was obtained from the Borexino data release corresponding to Ref.~\cite{Agostini:2018uly} and Ref~\cite{Bellini:2013lnn}. Our standard interactions (SI) prediction is shown in solid red. An example NSI solution that enhances the event spectra is shown in dashed red.}
    \end{center}
\end{figure}

\subsection{Atmospheric Neutrinos at DUNE}
Neutrinos produced in the Earth's atmosphere from cosmic ray processes consist of the $\nu_e$, $\bar{\nu}_e$, $\nu_\mu$, and $\bar{\nu}_\mu$ flavor states. These neutrinos are then free to propagate through the Earth and undergo flavor oscillations. The neutrinos can then be detected by the DUNE far detector, capable of reconstructing the neutrino energy and direction (or zenith angle between the incoming neutrino trajectory and the horizon plane at the detector). In particular, in Ref.~\cite{Kelly:2019itm} the atmospheric neutrino flux below $E_\nu = 1$ GeV has been shown to exhibit rich oscillation information, which not only aids in the measurement of the leptonic mixing angles and the CP violating phase $\delta_{CP}$, but also in gaining sensitivity to the NSI matter effects.

Since we limit the scope of this analysis to neutral-current vector NSI, we describe the charged-current interactions in LAr with the SM prediction, namely the processes
\begin{align}
    \nu_\alpha + n \to \ell^-_\alpha + p^+ \nonumber \\
    \bar{\nu}_\alpha + p^+ \to \ell^+_\alpha + n \nonumber
\end{align}
via charged-current quasi-elastic (CCQE) scattering. We ignore resonance production processes, which will reduce statistics but mitigates the theoretical uncertainties in the resonance production cross-sections as well as the hadronic energy corrections that smear the energy reconstruction~\cite{Friedland:2018vry}. Additionally, we will restrict ourselves to $\nu_\mu$ scattering, whose final state typically gives rise to two well-identified charged tracks in the detector.

For the analytic form of the cross-section $\sigma(E_\nu)$ we implement the one developed by in Ref.~\cite{Bodek:2011ps} which includes a parameterization of the transverse enhancement from meson exchange currents inside the nucleus. This parameterization offers a decent fit to cross-section data at in the relevant energy range for atmospheric neutrinos of 100 MeV to 1 GeV. We provide more details of the implementation of this cross-section in Appendix \ref{app:ccqe}.

Atmospheric fluxes $\Phi_\alpha (\cos\theta , E_\nu)$ are taken from the FLUKA results of Ref.~\cite{Battistoni:1999at} for the Super-Kamiokande site.
To obtain the predicted event count of a neutrino flavor $\alpha$ between energies $E_\nu^a$ and $E_\nu^b$ and zeniths $\cos\theta_1$ and $\cos\theta_2$, assuming perfect reconstruction, we convolute $\sigma(E_\nu)$ with the oscillated atmospheric flux. The number of neutrinos observed for a flavor $\alpha$ is then given in Eq.~\ref{eq:atmos_rate}:
\begin{equation}
N_\alpha = 2\pi \int_{E_{\nu}^a}^{E_{\nu}^b} \int_{\cos\theta_1}^{\cos\theta_2} \frac{\mathcal{E}}{M_T} \sigma(E_{\nu}) 
\bigg\{\sum_\beta \dfrac{\partial^2 \Phi_\beta}{\partial E_\nu \partial \Omega}
 P_{\beta\alpha} \bigg\}
 d E_\nu d(\cos\theta)
\label{eq:atmos_rate}
\end{equation}
$P_{\beta\alpha} = P_{\beta\alpha}(\cos\theta , E_\nu)$ are the survival and transition probabilities determined from the neutrino Hamiltonian with NSI. To calculate $P_{\beta\alpha}$ we employ a numerical diagonalization method on the oscillation hamiltonian in conjunction with a simplified version of the PREM model for the electron number density in the Earth.

\begin{figure*}[!tbh]
\centering
      \includegraphics[width=\textwidth]{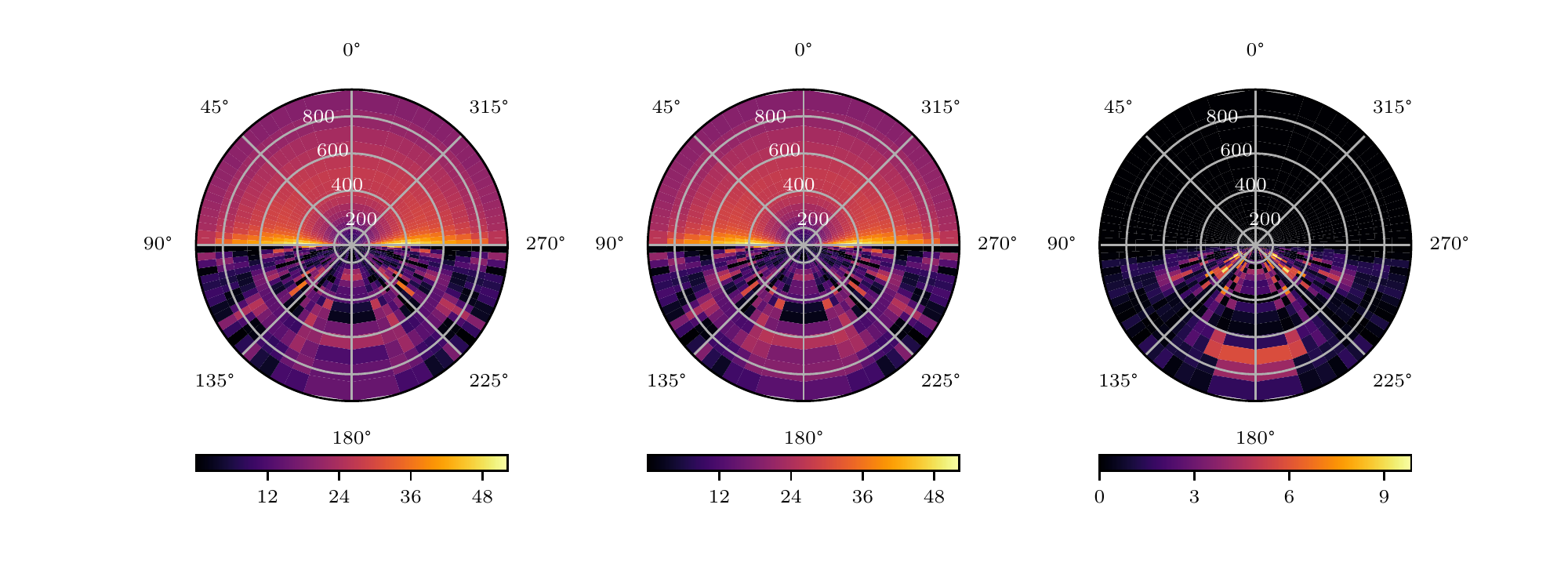}
       \caption{\label{fig:dune_rates}Polar plot of the DUNE atmospheric $\nu_\mu$ appearance rates over 40 zenith bins from $\cos\theta \in [-0.975, 0.975]$ ($\theta$ plotted here as the polar angle) and $E_\nu \in [100, 1000]\,$ MeV (plotted in the radial direction). Standard interactions ($\epsilon^O_{\alpha\beta} = 0$, left), NSI ($\epsilon^O_{ee} = 0.2$, $\epsilon^O_{\tau\tau} = 0.1$, $\epsilon^O_{e\tau} = 0.2$, center) and their difference $\mid N_{SI} - N_{NSI}\mid$ (right) are plotted.}
\end{figure*}

The rich spectrum is shown in Figure~\ref{fig:dune_rates} for all zenith angles and energies between 100 MeV and 1 GeV. The most NSI-sensitive region lies below the horizon, so we select 20 zenith bins for $\cos\theta \in [-0.975,-0.025]$. This corresponds to an angular resolution of about $18^\circ$. We use 20 energy bins between 100 MeV and 1 GeV. We take a 10 year exposure with the full 40 kton far detector volume. We then employ the same method as in Eq.~\ref{eq:loglike} but now defining the log-likelihood function over both energy and zenith bins. Since neutrino oscillations are sensitive to $\epsilon^{e,V}_{\alpha\beta}$, $\epsilon^{d,V}_{\alpha\beta}$, and $\epsilon^{d,V}_{\alpha\beta}$ in the Earth's matter potential, we allow all 18 real NSI degrees of freedom to vary in the likelihood function.
\\

\subsection{Solar and Atmospheric Neutrinos at a Future LXe Dark Matter Detector}
The atmospheric and solar neutrino event rate at a LXe dark matter detector via CE$\nu$NS (Eq.~\ref{eq:cevns}) or E$\nu$ES (Eq.~\ref{eq:eves}), respectively, can be predicted via a similar convolution to Eq.~\ref{eq:atmos_rate}. For the LXe detector we assume no direction reconstruction on the incoming neutrino and no flavor-sense; just energy reconstruction via nuclear recoils. Therefore we only use the zenith-integrated flux and sum over incoming neutrino flavors. For the statistical analysis of the predicted data, we again use a log-likelihood as in Eq.~\ref{eq:loglike}. Once again, for solar neutrinos we use a sum over energy bins and we include 12 NSI degrees of freedom (6 $\epsilon^{e,L}_{\alpha\beta}$ and 6 $\epsilon^{e,R}_{\alpha\beta}$) just as in the Borexino analysis. Since atmospheric neutrinos may oscillate into $\tau$ flavors in the Earth and interact in the detector via CE$\nu$NS, we are now sensitive to $\epsilon^{u,V}_{\tau\tau}$ and $\epsilon^{d,V}_{\tau\tau}$, thereby expanding our NSI degrees of freedom from the set used in the COHERENT analysis from 10 to 12 NSI.

\begin{figure}[!tbh]
  \begin{center}
  \includegraphics[width=0.6\textwidth]{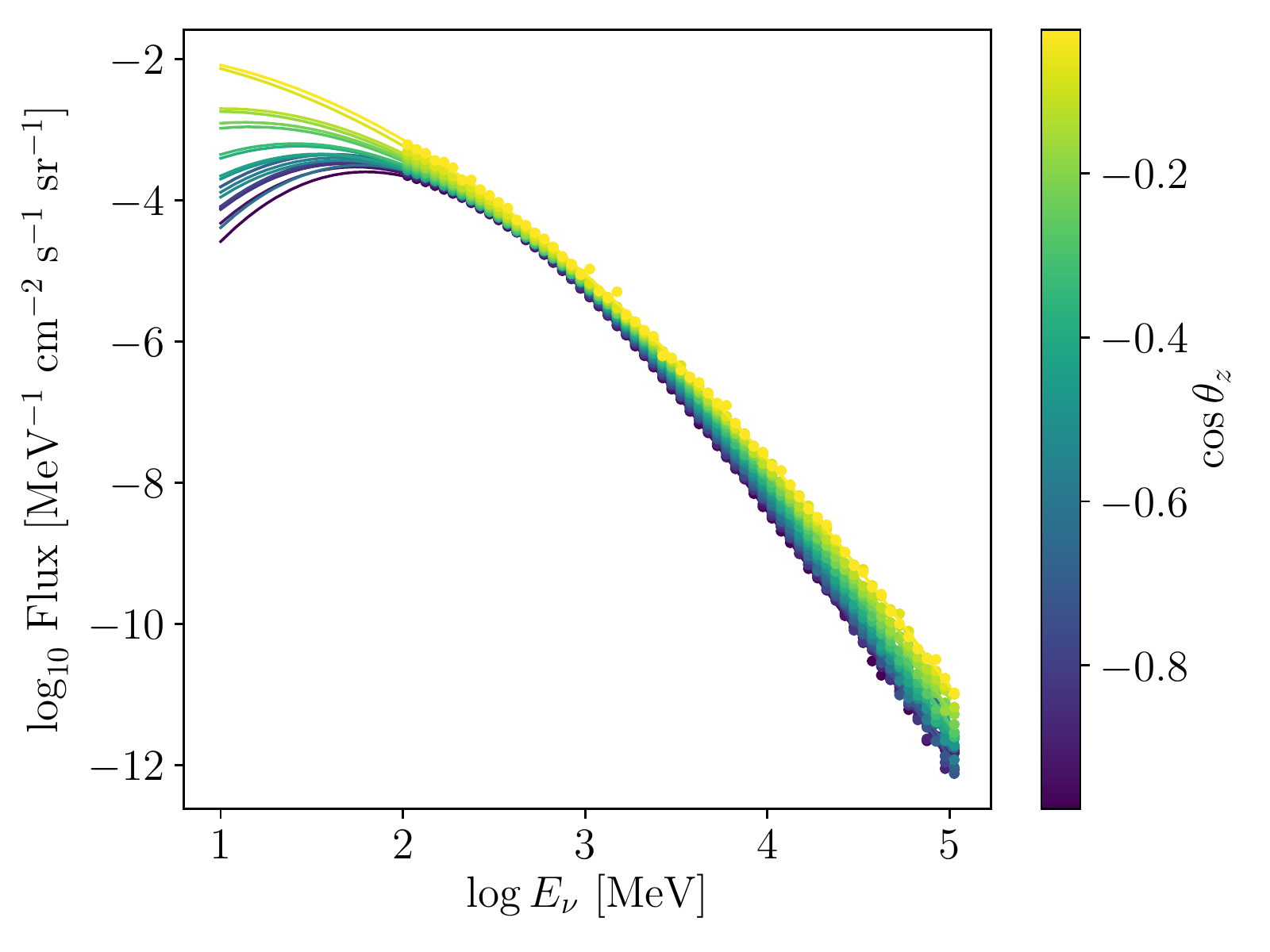}
    \caption{\label{fig:flux_extrap} Unoscillated atmospheric neutrino fluxes from FLUKA at the Super-Kamiokande site, extrapolated down to 10 MeV. We use a 3rd-order spline fit in log space in order to perform the extrapolation, i.e., $\log \Phi_\nu = \alpha + \beta \log E_\nu + \gamma (\log E_\nu)^2 + \delta (\log E_\nu)^3$ for fit constants $\alpha,\beta,\gamma,\delta$.}
    \end{center}
\end{figure}

We set the design goal exposure for this future detector to be 1 kton$\cdot$year. While this is larger than existing proposals in the literature~\cite{Aalbers:2016jon}, we take the approach of understanding the physics reach of such an experiment for an optimistic exposure. For the energy threshold, we assume recoils can be realistically reconstructed as low as 5 keV, looking for CE$\nu$NS events up to 50 keV. To see CE$\nu$NS events from atmospheric neutrinos, again we take the FLUKA result for the atmospheric flux, but an important point needs to be raised regarding this flux; atmospheric neutrinos need to have low enough energies ($\lesssim 60$ MeV) to scatter coherently off Xe nuclei, but the 3D FLUKA result only goes as low as 106 MeV. Therefore we extrapolate the atmospheric neutrino FLUKA fluxes down to 10 MeV using a 3rd-order polynomial in log space, shown in Figure~\ref{fig:flux_extrap}. While calculations of the zenith-by-zenith flux do not exist yet down to 10 MeV, we can check that the zenith-integrated flux agrees well with the one reported in in Ref.~\cite{Battistoni:2005pd} for the solar-averaged flux at Super-Kamiokande.

There is one final remark; this class of detector would also be sensitive to CE$\nu$NS from solar neutrinos coming from the $^8$B processes, inducing nuclear recoils up to energies of a few keV. For the analysis in this paper, we do not include this contribution to the event rate, in order to focus on atmospheric neutrino-induced events greater than 5 keV energy recoil. Extracting the recoils from $^8$B neutrinos would require a more dedicated analysis, as the complete detector efficiencies are difficult to estimate at this stage.

A quantitative summary of the specifications for each experiment simulated in this section can be found in Table~\ref{tab:experiments}.

\begin{table*}[t]
  \centering
  \caption{\label{tab:experiments} Kinematic configurations and exposures for each experiment considered.}
  \begin{tabular}{ l c c l l}
 Experiment & $E_\nu$ range (MeV) & $E_r$ range (keV)  & Cross-section & Exposure \\
 \hline
 Borexino (solar) & - & $[550,1000]$ & E$\nu$ES (NSI) & 92.1 kton$\cdot$days \\
 COHERENT (CsI)&- & $[5,23]$ & CE$\nu$NS (NSI) & 4466 kg$\cdot$days \\
 COHERENT (LAr)&- & $[20,100]$ &  CE$\nu$NS (NSI) & 15.33 ton$\cdot$days \\
 Future Xe (atmos.) & -& $[5,50]$ &  CE$\nu$NS (NSI)& 1 kton$\cdot$years \\
Future Xe (solar) & - & $[5,1000]$ &  E$\nu$ES (NSI)& 1 kton$\cdot$years \\
 DUNE (atmos.) & $[100,1000]$ & - & CCQE (SM)& 400 kton$\cdot$years
  \end{tabular}
\end{table*}

\section{Results}\label{sec:results}
\begin{figure*}[tbh]
  \begin{center}
   \includegraphics[width=\textwidth]{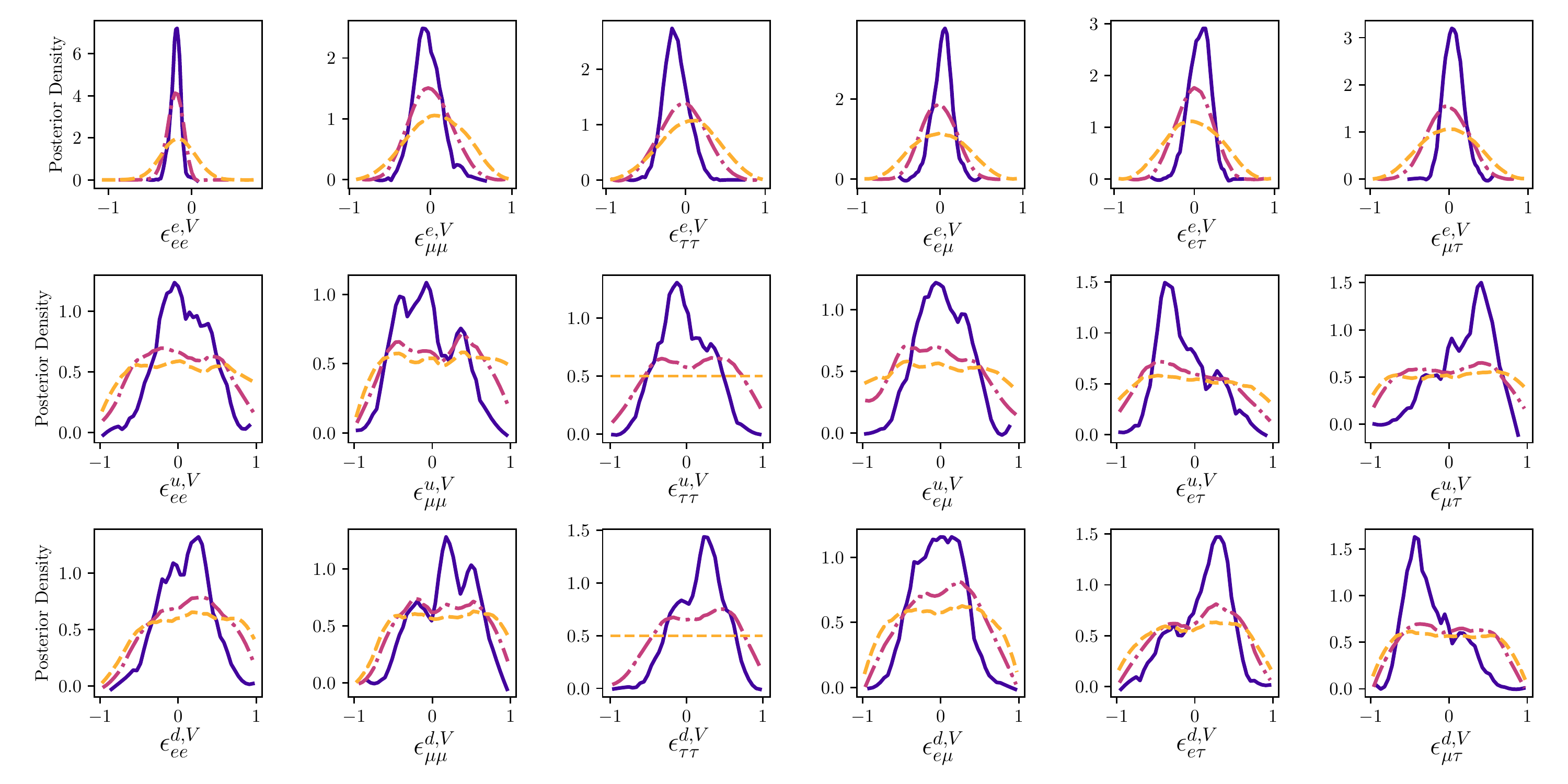}
    \caption{\label{fig:priorflow_marginals} Marginals of the posterior distributions in the prior flow as shown for the different stages in Figure~\ref{fig:priorflow}. The first stage (dashed yellow) consists of COHERENT and Borexino. The second stage (dash-dotted magenta) consists of the future LXe projection for atmospheric neutrinos and solar neutrinos and takes in priors from the first stage. The last stage (solid indigo) uses atmospheric neutrinos at DUNE and takes in priors from the second stage.}
    \end{center}
\end{figure*}

Using the methods we have just described, we derive the fits to the NSI parameters at each stage of the ``prior-flow" outlined in Figure~\ref{fig:priorflow}, with a final joint posterior distribution derived from the last stage at DUNE. We show the 1-dimensional marginalized posterior distributions for the 18 real-valued vector NSI parameters at various stages of the prior-flow in Figure~\ref{fig:priorflow_marginals}. In addition, the 95\% credible intervals corresponding to the 18 electron, $u$ and $d$ quark NSI are listed in Table~\ref{tab:credible}. Good convergence on the electron NSI $\epsilon^{e,V}_{\alpha\beta}$ is observed, but we note that the posterior means for $\epsilon^{e,V}_{ee}$ and $\epsilon^{e,V}_{\mu\mu}$ are slightly negative to accommodate the best-fit on the Borexino data about the $^7$Be edge. In the table we compare the credible intervals for COHERENT and Borexino (middle column) with those for the projections at DUNE and the LXe DMD. We observe that DUNE and the LXe DMD make an improved reduction in the width of the credible intervals on electron NSI by a factor of 2 to 3.

Convergence for $\epsilon^{u,V}_{\alpha\beta}$ and $\epsilon^{d,V}_{\alpha\beta}$ is also improved by DUNE and the LXe DMD with respect to the posteriors from COHERENT. COHERENT and the LXe DMD offer good constraints on the $u$ and $d$ quark NSI that enter in as priors for DUNE, but the constraints on electron NSI from solar neutrinos also help constrain the $u$ and $d$ quark NSI indirectly via the linear correlation $\epsilon^E_{ee} = \epsilon^{e,V}_{\alpha\beta} + 3\epsilon^{u,V}_{\alpha\beta} + 3\epsilon^{d,V}_{\alpha\beta}$, which enters into the matter potential to which DUNE is sensitive. Phenomenologically speaking, a strong constraint on $\epsilon^{e,V}_{\alpha\beta}$ incurs an equal and opposite constraint on $3\epsilon^{u,V}_{\alpha\beta} + 3\epsilon^{d,V}_{\alpha\beta}$. Note, however, that this relationship can also have the effect of inducing biases; if the data at DUNE is consistent with $\epsilon^E_{ee} = 0$, then via the aforementioned linear combination, any bias in $\epsilon^{e,V}_{ee}$ will induce a bias in $\epsilon^{u,V}_{\alpha\beta}$ and $\epsilon^{d,V}_{\alpha\beta}$ via their correlations through $\epsilon^E_{ee}$. We see this effect notably in $ee$ and $\mu\mu$ NSI caused by the negative bias in $\epsilon^{e,V}_{ee}$ from the Borexino part of the analysis. It should also be pointed out here that the biases seen in Figure~\ref{fig:priorflow_marginals} are reflected in the credible intervals in Table~\ref{tab:credible}; for some NSI, the fit has pushed the credible interval to exclude the zero value point, but again this arises as an artifact of the null hypotheses we have assumed for DUNE and the LXe DMD and the intrinsic correlation between the fits on $\epsilon^{e,V}_{ee}$, $\epsilon^{u,V}_{\alpha\beta}$, and $\epsilon^{d,V}_{\alpha\beta}$ NSI.

\begin{figure*}[tbh]
  \begin{center}
   \includegraphics[width=1.15\textwidth]{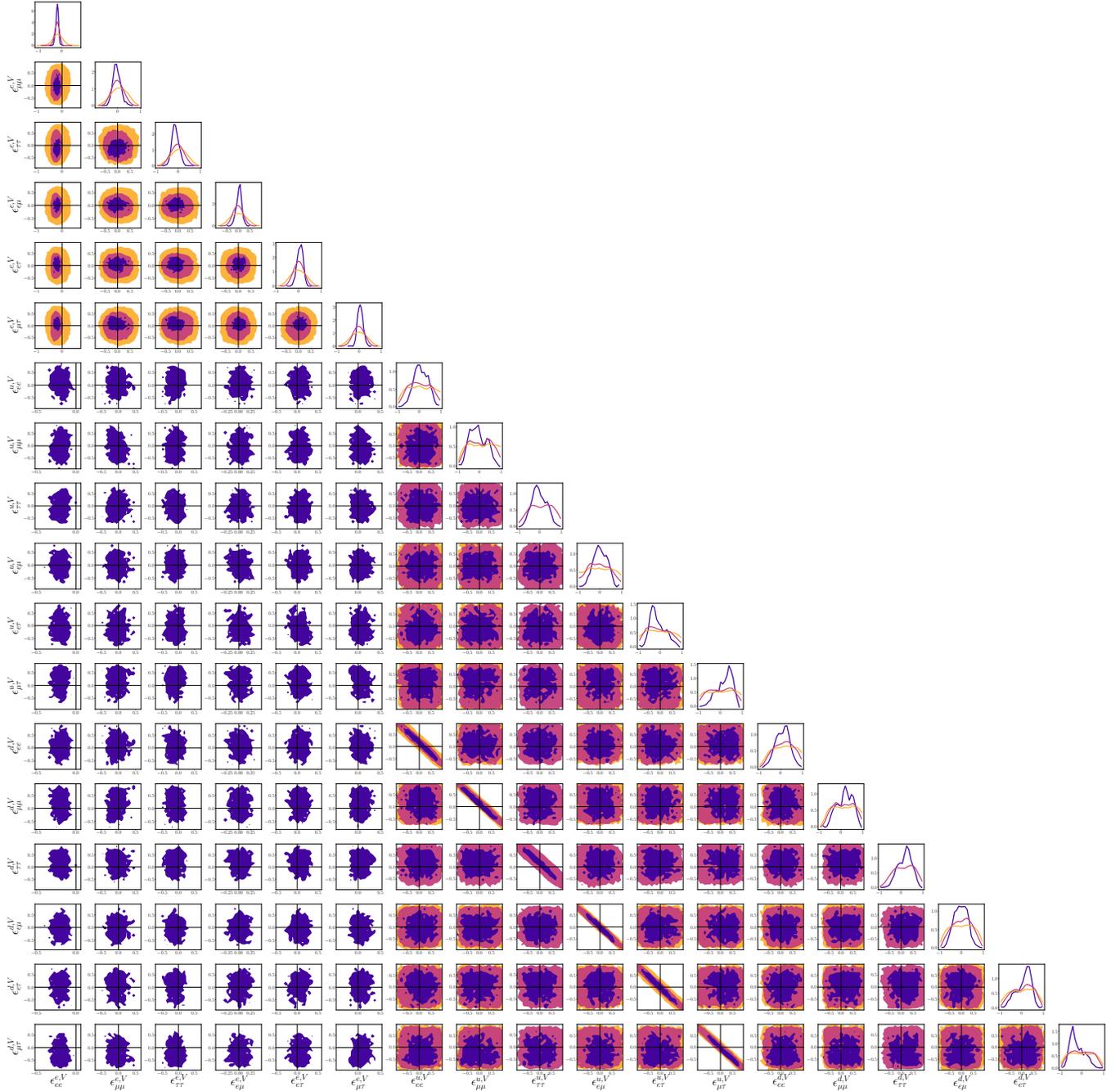}
    \caption{\label{fig:priorflow_18d} 1-dimensional marginals and 95\% credible contours for the 2-dimensional marginal projections are shown for $\epsilon^{e,V}_{\alpha\beta}$, $\epsilon^{u,V}_{\alpha\beta}$, and $\epsilon^{d,V}_{\alpha\beta}$ for a total of 18 NSI degrees of freedom. The first stage (yellow) consists of COHERENT and Borexino. The second stage (magenta) consists of the future LXe projection for atmospheric neutrinos and solar neutrinos and takes in priors from the first stage. The final result at the last stage (indigo) uses atmospheric neutrinos at DUNE and takes in priors from the second stage.}
    \end{center}
\end{figure*}

Additionally, even with multiple detector materials available in our analysis to break the $\epsilon^{u,V}_{\alpha\beta}$-$\epsilon^{d,V}_{\alpha\beta}$ degeneracy, some degeneracy still remains from the correlation between $u$ and $d$ quark NSI in the 2-dimensional marginal posterior distributions; see Figure~\ref{fig:priorflow_18d} where we show all the 2-dimensional projections of the prior-flow posteriors in 18 NSI dimensions. The credible regions in the $(\epsilon^{u,V}_{\alpha\beta}, \epsilon^{d,V}_{\alpha\beta})$ planes are certainly improving with each stage, but the correlation never fully goes away.

\begin{figure*}[tbh]
  \begin{center}
   \includegraphics[width=1.15\textwidth]{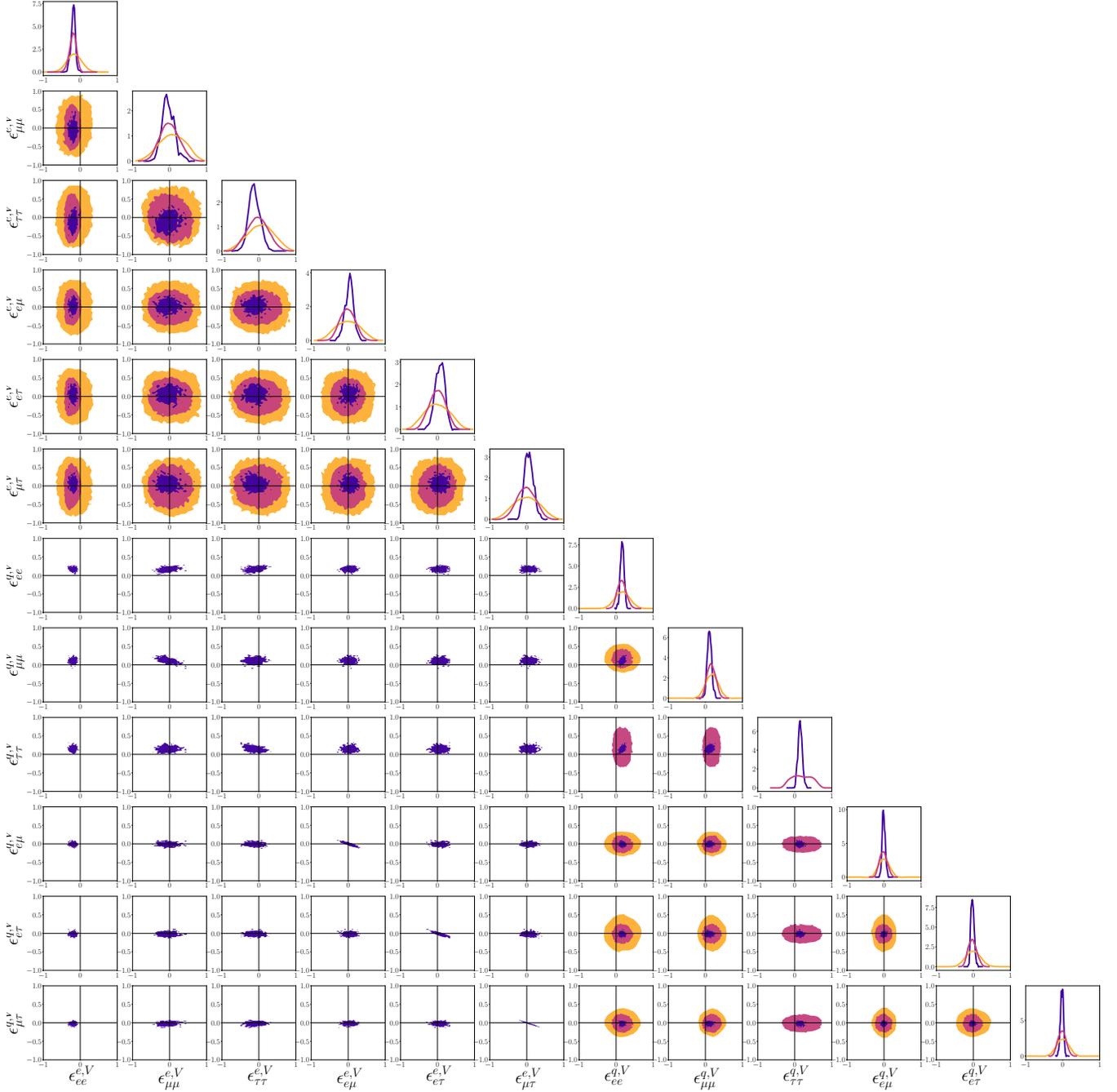}
    \caption{\label{fig:priorflow_12d} 1-dimensional marginals and 95\% credible contours for the 2-dimensional marginal projections are shown for each stage in the prior-flow, but now reduced to 12 NSI degrees of freedom by summing $\epsilon^{u,V}_{\alpha\beta} + \epsilon^{d,V}_{\alpha\beta} \equiv \epsilon^{q,V}_{\alpha\beta}$. The first stage (yellow) consists of COHERENT and Borexino. The second stage (magenta) consists of the future LXe projection for atmospheric neutrinos and solar neutrinos and takes in priors from the first stage. The final result at the last stage (indigo) uses atmospheric neutrinos at DUNE and takes in priors from the second stage.}
    \end{center}
\end{figure*}

By defining $\epsilon^{q,V}_{\alpha\beta} \equiv \epsilon^{u,V}_{\alpha\beta} + \epsilon^{d,V}_{\alpha\beta}$ we can transform away the strong correlation between $u$ and $d$ quark NSI and visualize the remaining degeneracy between electron and quark NSI that DUNE would exhibit. In Figure~\ref{fig:priorflow_12d} we plot a grid of the 1- and 2-dimensional marginal projections of the NSI parameters reduced to just 12 NSI (6 $\epsilon^{e,V}_{\alpha\beta}$ and 6 $\epsilon^{q,V}_{\alpha\beta}$). Again the double-solution degeneracy on $\epsilon^{q,V}_{\tau\tau}$ is broken by DUNE. After this transformation we observe good convergence on $\epsilon^{q,V}_{\alpha\beta}$ and $\epsilon^{e,V}_{\alpha\beta}$, with improved reduction in the credible interval widths by roughly a factor of 2 with the addition of DUNE and the LXe DMD as shown in Table~\ref{tab:credible}. Some correlation remains between the pairwise combinations of $\epsilon^{q,V}_{\alpha\beta}$ and $\epsilon^{e,V}_{\alpha\beta}$, most prominantly between $\epsilon^{q,V}_{\mu\tau}$ and $\epsilon^{e,V}_{\mu\tau}$. Overall, the reduced set of 12 NSI comprising $\epsilon^{e,V}_{\alpha\beta}$ and $\epsilon^{q,V}_{\alpha\beta}$ has the best convergence with the most number of degeneracies broken, while still representing a set of NSI parameters that are not too phenomenological to be non-influential to model-building.

\begin{figure*}[tbh]
  \begin{center}
   \includegraphics[width=1.15\textwidth]{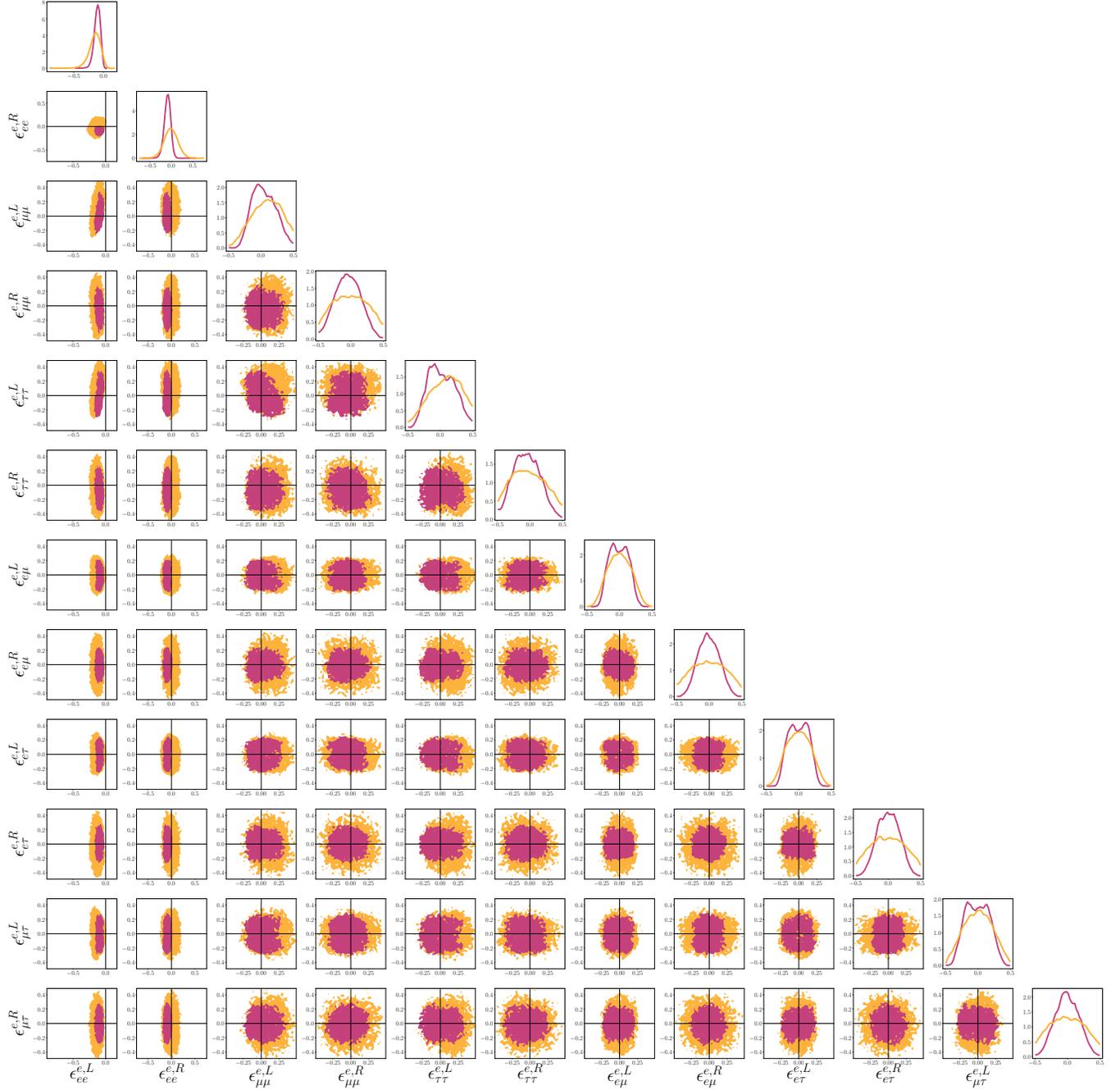}
    \caption{\label{fig:priorflow_lr} 1-dimensional marginals and 68\% credible contours for the 2-dimensional marginal projections are shown for left-handed and right-handed $e$-NSI ($\epsilon^{e,L}_{\alpha\beta}$, $\epsilon^{e,R}_{\alpha\beta}$) for a total of 12 NSI degrees of freedom. The distributions and contours for Borexino (yellow) and a future LXe DMD with priors from Borexino (magenta) are overlayed.}
    \end{center}
\end{figure*}

Finally, we also show in Figure~\ref{fig:priorflow_lr} the posteriors for $\epsilon^{e,L}_{\alpha\beta}$ and $\epsilon^{e,R}_{\alpha\beta}$ NSI before they are passed in as priors for DUNE in their vector combinations. The 68\% credible contours and 1-dimensional marginal posterior probability distributions are compared between Borexino and a future LXe DMD. Excellent convergence is achieved on $\epsilon^{e,L}_{ee}$ and $\epsilon^{e,R}_{ee}$ due to the CC enhancement to the E$\nu$ES cross-section which constructively interferes with $\epsilon^{e,L}_{ee}$ and $\epsilon^{e,R}_{ee}$ NSI to produce larger effects on the $^7$Be flux.

\begin{table*}[tbh]
  \centering
  \caption{\label{tab:credible} We show 95\% credible intervals for the NSI parameters derived from existing data from Borexino and COHERENT (middle column), and projected constraints from the combined results of a LXe DM detector and DUNE (right column) whose priors were constructed from the posterior distributions on NSI from COHERENT and Borexino.}
\begin{tabular}{ | c | c | c | } 
\hline
NSI & \shortstack{Borexino,\\COHERENT (CsI \& LAr)} & \shortstack{+LXe DM Detector,\\DUNE (Future)} \\ 
\hline
$\epsilon^{e,V}_{ee}$ & $[-0.56,0.24]$ & $[-0.31,-0.084]$ \\ 
\hline
$\epsilon^{e,V}_{\mu\mu}$ & $[-0.58,0.72]$ & $[-0.35,0.32]$ \\ 
\hline
$\epsilon^{e,V}_{\tau\tau}$ & $[-0.60,0.72]$ & $[-0.35,0.20]$ \\ 
\hline
$\epsilon^{e,V}_{e\mu}$ & $[-0.58,0.60]$ & $[-0.21,0.25]$ \\ 
\hline
$\epsilon^{e,V}_{e\tau}$ & $[-0.60,0.62]$ & $[-0.18,0.31]$ \\ 
\hline
$\epsilon^{e,V}_{\mu\tau}$ & $[-0.67,0.62]$ & $[-0.18,0.28]$ \\ 
\hline
$\epsilon^{u,V}_{ee}$ & $[-0.88,0.98]$ & $[-0.53,0.72]$ \\ 
\hline
$\epsilon^{u,V}_{\mu\mu}$ & $[-0.82,0.97]$ & $[-0.71,0.61]$ \\ 
\hline
$\epsilon^{u,V}_{\tau\tau}$ & - & $[-0.67,0.62]$ \\ 
\hline
$\epsilon^{u,V}_{e\mu}$ & $[-0.92,0.98]$ & $[-0.53,0.55]$ \\ 
\hline
$\epsilon^{u,V}_{e\tau}$ & $[-0.97,0.87]$ & $[-0.72,0.62]$ \\ 
\hline
$\epsilon^{u,V}_{\mu\tau}$ & $[-0.92,0.92]$ & $[-0.48,0.64]$ \\ 
\hline
$\epsilon^{d,V}_{ee}$ & $[-0.67,0.97]$ & $[-0.57,0.63]$ \\ 
\hline
$\epsilon^{d,V}_{\mu\mu}$ & $[-0.68,0.97]$ & $[-0.42,0.77]$ \\ 
\hline
$\epsilon^{d,V}_{\tau\tau}$ & - & $[-0.46,0.67]$ \\ 
\hline
$\epsilon^{d,V}_{e\mu}$ & $[-0.82,0.87]$ & $[-0.58,0.52]$ \\ 
\hline
$\epsilon^{d,V}_{e\tau}$ & $[-0.87,0.87]$ & $[-0.6,0.71]$ \\ 
\hline
$\epsilon^{d,V}_{\mu\tau}$ & $[-0.92,0.82]$ & $[-0.73,0.43]$ \\ 
\hline
$\epsilon^{q,V}_{ee} = \epsilon^{u,V}_{ee} + \epsilon^{d,V}_{ee}$ & $[-0.20,0.46]$ & $[0.070,0.26]$ \\ 
\hline
$\epsilon^{q,V}_{\mu\mu}= \epsilon^{u,V}_{\mu\mu} + \epsilon^{d,V}_{\mu\mu}$ & $[-0.049,0.46]$ & $[-0.035,0.24]$ \\ 
\hline
$\epsilon^{q,V}_{\tau\tau}=\epsilon^{u,V}_{\tau\tau}+\epsilon^{d,V}_{\tau\tau}$ & - & $[0.036,0.26]$ \\ 
\hline
$\epsilon^{q,V}_{e\mu}=\epsilon^{u,V}_{e\mu}+\epsilon^{d,V}_{e\mu}$ & $[-0.17,0.21]$ & $[-0.083,0.077]$ \\ 
\hline
$\epsilon^{q,V}_{e\tau}=\epsilon^{u,V}_{e\tau}+\epsilon^{d,V}_{e\tau}$ & $[-0.33,0.31]$ & $[-0.090,0.11]$ \\ 
\hline
$\epsilon^{q,V}_{\mu\tau}=\epsilon^{u,V}_{\mu\tau}+\epsilon^{d,V}_{\mu\tau}$ & $[-0.19,0.32]$ & $[-0.11,0.070]$ \\
\hline
  \end{tabular}
\end{table*}

\clearpage
\section{Conclusion}\label{sec:conclusion}
We have shown that it is possible to measure neutrino NSI, significantly breaking their degeneracies, even when many NSI parameters are nonzero. The inclusion of three different classes of observables - the CE$\nu$NS and E$\nu$ES processes and neutrino oscillations - are essential to constructing a global analysis whose experimental data are complementary to one another in the NSI model parameter space. We have chosen COHERENT and Borexino data sets as excellent representative neutrino scattering data sets, but these can readily be augmented with a variety of others. The far detector at DUNE with its large volume should provide excellent constraints on NSI through its ability to access rich oscillation information through the detection of atmospheric neutrinos after they interact with the matter potential of the Earth. The addition to this ensemble of neutrino scattering data at future dark matter experiments we showed to be a natural complement to the CE$\nu$NS data at COHERENT by their potential sensitivity to $\tau$ flavor neutrinos from solar and atmospheric sources. We stress that the experiments considered here are best used together as a unified source of data to investigate neutrino NSI.

The relatively many NSI considered in the analysis and multiple experiments being simulated became pragmatically realizable with our divide-and-conquer approach using the copula. We demonstrated that a strategy of connecting posterior probability distributions as Bayesian priors from experiment to experiment allows one to scale a global analysis with a potentially large number of model and nuisance parameters, with copulas facilitating the transfer of prior information. This novel ``prior-flow" framework we outlined can be extended in a straightforward way to include other existing data which would be sensitive to NSI. The Bayesian estimation of posterior probability distributions on the relatively large number of NSI parameters considered here was demonstrated to be tractable.

Our analysis could be extended, notably, to include neutrino-nucleus scattering data from CHARM, whose measurement of the cross-section ratio of NC to CC processes provides a well-known complementary constraint to CE$\nu$NS measurements in the $(\epsilon^{u,V}_{\alpha\beta}, \epsilon^{d,V}_{\alpha\beta})$ plane. It was omitted from this work for not providing a strong enough constraint relative to the parameter ranges we restricted ourselves to ($\epsilon^{f,V}_{\alpha\beta}\in[-1,1]$), but for a broader parameter space it would be interesting to integrate CHARM data into the analysis strategy~\cite{Coloma:2017egw}. Additionally, there are numerous oscillation data sets readily available which could contribute to the statistical power of the analysis, integrated into a global NSI study in a similar manner to DUNE.

To further generalize the projected constraints on NSI, we plan to extend our investigation to include complex-valued NSI parameters as well as effective NSI operators in scenarios where the underlying mediator masses are light and comparable to the scale of the neutrino momentum transfer. We emphasize the importance to obtain constraints on the NSI in these more general scenarios in order to support model-independent results and drive more theoretical work in this area. Since the space of neutrino experiments is expanding quickly and allowing for highly comprehensive analyses in the future, the need for new tools to combat model parameter degeneracies in highly generalized settings will be highly sought after. It is precisely these degeneracies that should make the reader appreciate that neutrino scattering and oscillation experiments should be thought of together, as a unified source of experimental information on new physics. We hope to have cut a pathway with the unique strategy presented here to give global analyses of neutrino interactions the ability scale up as we enter the precision frontier of neutrino physics.

\section*{Acknowledgements}
We thank S. Agarwalla, P. Denton, and A. Formozov for their helpful discussions. BD, SL, LES and AT acknowledge support from DOE grant DE-SC0010813. RFL acknowledges support from the National Science Foundation through grant PHY1719271. AT and SL also thank the Mitchell Institute for Fundamental Physics and Astronomy for Support. The authors acknowledge the Texas A\&M University Brazos and Terra HPC clusters that contributed to the research reported here.

\appendix
\section{CCQE Cross-section}\label{app:ccqe}
We will now review the charged current quasi-elastic (CCQE) scattering process in liquid $^{40}_{18}$Ar. This concerns the reactions $\nu_\alpha + n \to \ell^-_\alpha + p^+$ and $\bar{\nu}_\alpha + p^+ \to \ell^+_\alpha + n$ taking place with the protons and neutrons in the nucleus. For the cross-section and form factors we refer to Refs.~\cite{Bodek:2007ym,Day:2012gb}. Ref.~\cite{Formaggio:2013kya} also provides a very comprehensive review but note that equation 57 has incorrectly flipped the sign assignment for $\nu$ and $\bar{\nu}$ scattering cases.

The CCQE differential cross-section as a function of the momentum transfer $Q$ is
\begin{equation}
\label{eq:ccqe}
\dfrac{d\sigma(E_\nu, Q^2)}{dQ^2} = \dfrac{M^2 G_F^2 \cos^2\theta_c}{8\pi E_\nu^2} \bigg[A(Q^2) \pm B(Q^2)\dfrac{s-u}{M^2}
+ C(Q^2)\dfrac{(s-u)^2}{M^4} \bigg]
\end{equation}
where the $+$ sign is taken for $\Bar{\nu}$ and $-$ for $\nu$ scattering. $E_\nu$ is the energy of the initial state neutrino, $M$ is the target nucleon mass, and $s-u = 4ME_\nu - Q^2 - m^2$ where $m$ is the mass of the final state lepton. Let $\tau \equiv Q^2/4M^2$ and let $\xi \equiv \mu_p - \mu_n = 4.706$, with $m$ being the outgoing lepton mass and $M$ the target nucleon mass (neutron or proton). We take the global fit to the axial mass $M_A = 1014$ MeV. The $A$, $B$, and $C$ terms are given as follows
\begin{align}
    A(Q^2) = \dfrac{m^2 + Q^2}{M^2} \bigg[ &(1+ \tau)|F_A|^2 - (1-\tau)|F_V^2|^2 - \tau (1 - \tau) |F_V^2|^2 - 4\tau F_V^1 F_V^2  \nonumber \\
     &- \frac{m^2}{4M^2}\bigg(|F_V^1 +  F_V^2|^2 + |F_A + 2F_p|^2 - 4(1+\tau)|F_p|^2 \bigg) \bigg] 
\end{align}
\begin{equation}
    B(Q^2) = 4 \tau F_A (F_V^1 + F_V^2)
\end{equation}
\begin{equation}
    C(Q^2) = \frac{1}{4} \bigg(|F_A|^2 + |F_V^1|^2 + \tau|F_V^2|^2 \bigg)
\end{equation}

Each of the form factors above can be constructed from dipole terms;
\begin{equation}
    G_D \equiv \dfrac{1}{1 + \frac{Q^2}{M_V^2}},
\end{equation}
and we will take $G_E = G_D$ and $G_M = \xi G_D$. The transverse enhancement from meson exchange currents~\cite{Bodek:2007ym} is parameterized by
\begin{equation}
    \Theta = \sqrt{1 + a Q^2 \, e^{\frac{-Q^2}{b}}}
\end{equation}
where the best fit parameters are $a=6\times10^{-6} \,$ MeV$^{-2}$, $b = 3.5\times10^5$ MeV$^2$. The form factors can now be given.
 \begin{equation}
     F_1 = \dfrac{G_E + \tau \Theta G_M}{1 + \tau} = G_D\dfrac{1 + \tau\Theta\xi}{1+\tau}
 \end{equation}
\begin{equation}
    F_2 = \dfrac{\Theta G_M - G_E}{1 + \tau} = G_D \dfrac{\Theta-\xi}{1+\tau}
\end{equation}
 \begin{equation}
 F_A = \dfrac{-1.267}{(1 + \frac{Q^2}{M_A^2})^2}
 \end{equation}
 \begin{equation}
 F_p = \dfrac{2M^2 F_A (Q^2)}{M_\pi^2 + Q^2}
 \end{equation}

\begin{figure}[!tbh]
  \begin{center}
   \includegraphics[width=0.8\textwidth]{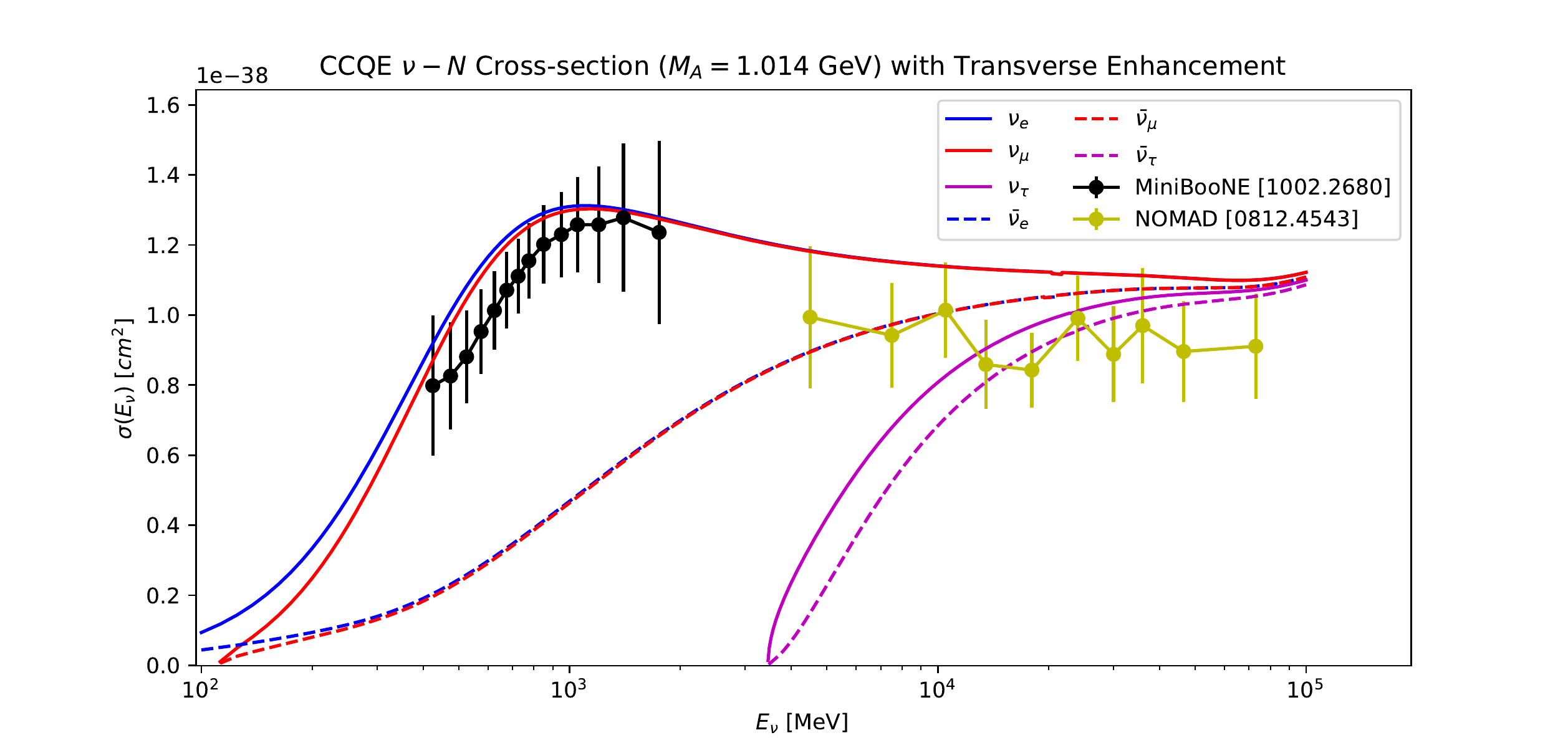}
    \caption{\label{fig:ccqe_all} The charged current quasi-elastic (CCQE) cross-section per nucleon is plotted by integrating Eq.~\ref{eq:ccqe} for each neutrino species. Only the $\nu_\mu$ scattering cross-section is used in this work to predict $\nu_\mu$ scattering rates at DUNE, for energies between 100 and 1000 MeV. The NOMAD~\cite{Lyubushkin:2008pe} and MiniBooNE~\cite{AguilarArevalo:2010zc} measurements of the $\nu_\mu$ cross-section are overlayed.}
    \end{center}
\end{figure}

To perform the $Q^2$ integration, we use
\begin{equation}
\sigma(E_\nu) = \int_{Q^2_{min}}^{Q^2_{max}} \dfrac{d\sigma(E_\nu , Q^2)}{dQ^2} dQ^2
\end{equation}
with $Q^2_{\substack{max\\min}} = - m_\ell^2 + \dfrac{s - M^2}{\sqrt{s}} (E_\ell \pm |p_\ell|)$, $E_l = \dfrac{s + m_\ell^2 - M^2}{2\sqrt{s}}$, $|p_\ell| = \sqrt{E_\ell^2 - m_\ell^2}$, and $\sqrt{s} = \sqrt{M^2 + 2ME_\nu}$.
Finally, we scale the total cross-section by the number of target nucleons in the $^{40}_{18}$Ar nucleus; by $Z = 18$ for $\bar{\nu}$ and by $N = 22$ for $\nu$ scattering. The cross-section per nucleon is plotted for each neutrino type and compared with NOMAD~\cite{Lyubushkin:2008pe} and MiniBooNE~\cite{AguilarArevalo:2010zc} data in Figure~\ref{fig:ccqe_all}.

\section{More on E$\nu$ES Degeneracies}\label{app:degen}
\begin{figure}[h]
 \centering
 \includegraphics[width=0.5\textwidth]{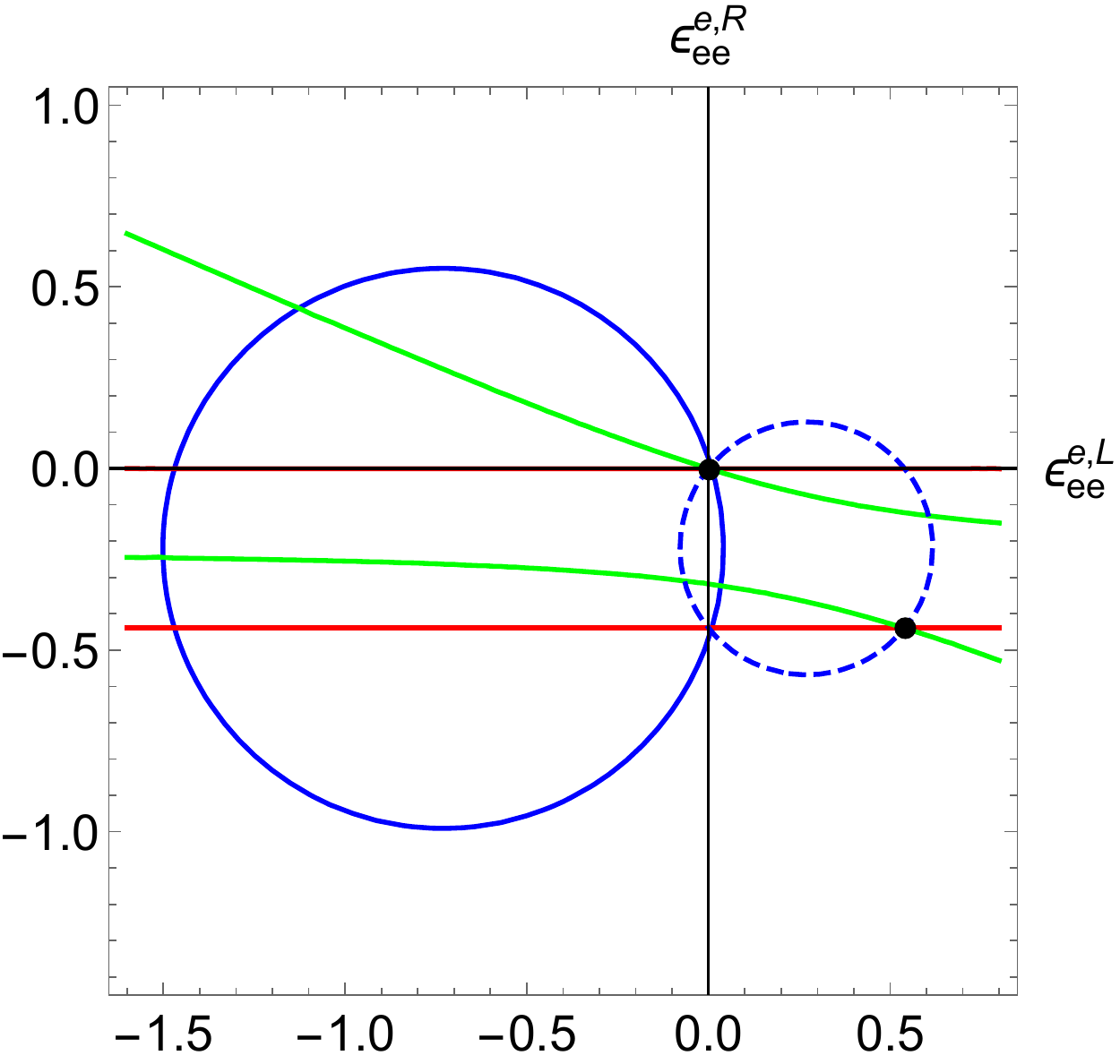}
\caption{\label{fig:intersecting_sols} SM degeneracy curves for each term in the $E_r$ expansion in the E$\nu$ES cross-section are shown. The black points indicate where the sets of curves intersect, corresponding to the NSI solutions for $\epsilon^{e,L}_{\alpha\beta}$ and $\epsilon^{e,R}_{\alpha\beta}$ that leave the E$\nu$ES cross-section degenerate with the SM.}
\end{figure}

In Figure~\ref{fig:intersecting_sols} a visualization of the degeneracy structure between $\epsilon^{e,L}_{\alpha\beta}$ and $\epsilon^{e,R}_{\alpha\beta}$ is shown. The curves shown are defined by equating the constant term (blue), the $E_r$ terms (green), and the $E_r^2$ terms (red) in the E$\nu$ES cross-section with their SM forms. In the case that the initial state neutrino is of electron flavor, the charged current enhancement leads to the solid blue circle which only intersects the other two curves once at the black point at the origin. For other initial state flavors, we obtain the dashed blue circle which intersects the other curves at both black points - a two-solution degeneracy with the SM.

\bibliography{main}
\bibliographystyle{JHEP}

\end{document}